# Impedance Spectroscopy for Electroceramics and Electrochemical System

Dr. Subrata Karmakar[1&2,*]

[1]Department of Physics & Astronomy, National Institute of Technology, Rourkela, 769008, India

[2]Electrical Engineering, Ingram School of Engineering, Texas State University, San Marcos, TX 78666, USA.

**Correspondence:** skarmakar@txstate.edu*,

## Abstract

This tutorial review focuses on the basic theoretical backgrounds, their working principles, and implementation of impedance spectroscopy in both electroceramics and electrochemical research and technological applications. Various contributions to the impedance, admittance, dielectric, and conductivity characteristics of electroceramic materials can be disentangled and independently characterized with the help of impedance spectroscopy as a function of frequency and temperature. In polycrystalline materials, the impedance, charge transport/ conduction mechanism, and the macroscopic dielectric properties i.e., dielectric constant and loss are typically composed of many contributions, including the bulk or grain resistance/capacitance, grain boundary, and sample-electrode interface effect. Similarly, electrochemical impedance spectroscopy (EIS) endeavors to the charging kinetics, diffusion, and mechanical impact of various electrochemical systems widely used in energy storage (i.e., supercapacitor, battery), corrosion resistance, chemical and bio-sensing, diagnostics, etc. in electrolytes as a function of frequency. The understanding of various contributions in the EIS spectra i.e., kinetic control, mass control, and diffusion control is essential for their practical implications. It is demonstrated that electrochemical and electroceramics impedance spectroscopy is an effective method to explain and simulate such behavior. Deconvolute these contributions to obtain a detailed understanding of the functionality of polycrystalline electroceramic materials. This short review aims to endow the expertise of senior researchers in many fields where both EIS (electrochemical and ceramics) are involved, as well as to provide the necessary background information for junior researchers working in these fields.

## 1. Introduction

In solid-state ceramic or bulk materials, the measurement of electrical parameters i.e., impedance, admittance, conductance, or dielectric constant as a function of frequency is known as impedance spectroscopy (IS). It is employed to resolve various electrical polarization processes based on their time constants or relaxation frequencies and this method can be used for dielectric materials, and ionic, or electronic conductors. It is frequently used in integrated energy conversion devices like supercapacitors, batteries, and fuel cells, as well as electroceramics, solid electrolytes, and dielectrics, including polymers and glasses. It is one of the primary methods used in electrochemistry to investigate the electrode processes and is often known as electrochemical impedance spectroscopy (EIS). The impedance spectroscopy is based on the application of a sinusoidal signal (ac voltage or current) over a wide range of frequencies ($\Omega$ to M$\Omega$) to perturb an electrical or electrochemical system in equilibrium or steady state, and the observation of the sinusoidal response (current or voltage, respectively) of the system to the externally applied perturbation.[1] The electroceramic or electrochemical system under study is a linear time-invariant system which means the behavior of the system is not affected by time over a wide range of frequencies and the output signal (ac current or ac voltage) is controlled and linearly related to the input signal (ac voltage or ac current) through a "transfer function" technique of IS. By using the IS, it is possible to identify, and frequently isolate reactions brought on by the bulk of the materials or devices, the existence of internal and external interfaces, and the acquisition of information about charge storage and dissipation.[2] These days, it is easy to find equipment that can measure frequencies over 1 to 12 orders of magnitude. This technique has found applications in the characterization of conventional capacitors, fuel cells, batteries, and solar cells because these devices have a wide range of electrical responses and multiple material characteristics.

The impedance spectroscopy has been utilized to monitor the aging of structural and electrical components as well as to control the quality of solid-state capacitors. Therefore, the technique is not exclusive to energy-related devices and gadgets. Regardless of the potential use, any material or device can be described to ascertain the contributions of the various components grain, grain boundary, and/or interfaces encountered, and their frequency relationships demonstrated at room or various temperatures. High-conducting materials are best represented in terms of conductivity as a function of temperature and/or working environment (vacuum or pressure), whereas responses for dielectric-type materials are best interpreted in the dielectric formalism, or sometimes both. Depending on the kind of material under investigation, several electrical processes can be solved using the IS. Electrical polarizations in polycrystalline materials occur due to the reduced electrical conductivity of grain boundaries in comparison to the interiors of the grains. There are three possible reasons for the decreased conductivity: second phases at the grain boundaries, space charges, or dopant segregation. IS has made a significant contribution to our understanding of ceramics used in electrochemical devices like solid oxide fuel cells. In the field of electroceramics, grain boundaries play a crucial role in determining the nonlinear current-voltage characteristics of varistors and the positive temperature coefficient of resistance (PTCR). Also, the electrical resistance of materials decreases as temperature rises and it is known as a negative temperature coefficient of resistance (NTCR).[3,4] This behavior defies the observation of most conductive materials, in which a positive temperature coefficient usually indicates an increase in resistance with temperature and the temperature dependency of charge carrier mobility or concentration is the main cause of the phenomenon in NTC materials. The impedance is inversely related to the dielectric permittivity/conductivity, and we get a lot of information about dielectric relaxation, conduction mechanism, etc. The reader is directed to more in-depth research articles where information is frequently concealed inside the incompletely analyzed impedance data and it can be interpreted by converting impedance to three additional formalisms known as conductivity, electric modulus, and permittivity.[5]

The ability of EIS spectroscopy to discriminate, and consequently to give a plethora of information for diverse electrical, electrochemical, and physical processes that take place in a real electrochemical system, sets it apart from other electrochemical techniques. This work is quite difficult because each of these processes has a unique time behavior ranging from extremely quick to extremely sluggish. Different time constants (τ) are displayed by various electrochemical processes and these are dependent on various factors such as the kinetics of an electrode charge-transfer reaction, the resistance of a liquid electrolyte, the distinct bulk/grain and grain boundary resistances for a solid polycrystalline electrolyte, the charging or discharging of the electric double layer at the electrode/electrolyte interface, the dependence of the electric double layer capacitive (EDLC) behavior on the morphology of the electrode surface and the composition of the electrolyte, the electrode charge-transfer reaction linked to the homogeneous reactions and adsorption/desorption phenomena, and diffusion controlled mass transfer phenomena, etc. The time constant (τ) of any electrochemical process can be represented by,[4,6]

$$\tau = RC = \frac{1}{2\pi v_p} \quad \ldots (1)$$

where $R$, $C$, and $v_p$ are the resistance, capacitance, and peak frequency of the Nyquist plots, respectively. It is noteworthy that IS measurements for an electrochemical or electroceramic system are usually correlated to an equivalent electrical circuit model consisting of three common passive components like resistors (R), capacitors (C), and inductors (L) and their multiple combinations which makes more complex (referred to as distributed) elements connected in various ways. Thus, each of these procedures may be considered analogous to a corresponding electrical circuit with a distinct time constant (τ). Most electrochemical analyzers come with the necessary software to simulate the impedance data to a model circuit or can be fitted in some other specialized analogous circuit modeling software i.e., ZSimpWin, Zview, or Zplot (Scribner Associates, Inc.).

In this review, we aim to focus on the basic principle of impedance spectroscopy, the strategy of calculations of various parameters, their implementations, and possible applications in real-life systems and research. Various review papers are available in the market that mainly focus on electrochemical impedance spectroscopy rather than solid-state electroceramic, bulk, nano, and polymer composite systems. This review accommodates information on impedance spectroscopy for both systems in a single platform and the conversion of impedance data to other different formalisms, known as admittance, electric modulus, dielectric constant, loss, conductivity, as well as the evolution of impedance with various conditions/environments in electrochemistry. Even though there is not enough space to go over all the subtleties of this technique in this short review, still we have managed to supply the basics of the conversions that are required to obtain the best fit and capture all the electrical parameters in a given experiment for the junior researchers or beginner.

2. **Basic Principle of Impedance Spectroscopy**

The basic principle behind an impedance spectroscopy experiment is to apply a sinusoidal electrical voltage or current, (depending on the sample) and track the output current or voltage response with frequency at different conditions i.e., temperatures or bias voltages. The resistance of any Ohmic cell is defined as the ratio of voltage to the current. Typically, an AC sinusoidal voltage is applied to an electrochemical cell, and the current flowing through the cell is measured to determine the electrochemical impedance (Z).[7,8]

$$Z = \frac{V(t)}{I(t)} \ldots (2)$$

Usually, the AC voltage ($V$) is applied in the input and observed in the output AC current or vice-versa to get the impedance value. The sinusoidal alternating current and voltages are represented in **Figure 1**a.

The input ac voltage which is applied to the system can be defined as,

$$V(t) = V_0 \sin(\omega t) \ldots (3)$$

and the resulting alternating current in the output with the same angular frequency ($\omega$) and additional phase ($\varphi$) is,

$$I(t) = I_0 \sin(\omega t + \varphi) \ldots (4)$$

The ac impedance at this frequency is measured by,

$$Z(\omega) = Z' + iZ'' = \frac{V_0 \sin(\omega t)}{I_0 \sin(\omega t + \varphi)} = \frac{V_0}{I_0} \exp(i\varphi) = |Z|(\cos\varphi + i\sin\varphi)$$

Comparing the real and imaginary parts of impedance,

$$Z' = |Z|\cos\varphi \ldots (5)$$

$$Z'' = |Z|\sin\varphi \ldots (6)$$

where, $|Z| = \sqrt{Z'^2 + Z''^2}$, and $\varphi = \tan^{-1}(\frac{Z''}{Z'})$

The "Lissajous Figure" shown in **Figure 1**b, is produced if the applied sinusoidal voltage is plotted on the X-axis and the sinusoidal response signal (I) is plotted on the Y-axis. Prior to the development of contemporary impedance spectroscopy instruments, the sole method for measuring impedance was Lissajous analysis.[9,10] The Nyquist plot is the representation of real and imaginary parts of impedance with a time constant and one typical plot is shown in **Figure 1**c.

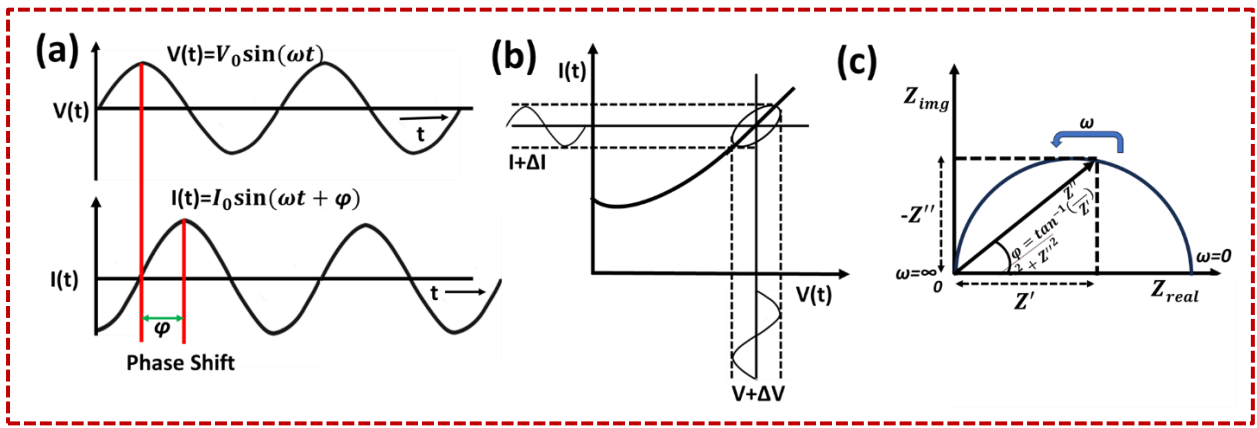

*Figure 1*. *(a) Sinusoidal current and voltage response in a linear electrical system, (b) the origin of Lissajous Figure, (c) complex impedance vectors and typical Nyquist plots of impedance spectroscopy.*

In any electrical (mostly dielectric system) or electrochemical system, one of the best ways to examine its electrical signal is by applying a sinusoidal input signal *x(t)* and observing the output signal *y(t)* which is shown in **Figure 2**. There are two different kinds of impedance spectroscopy widely used in research and technology-(i) solid-state impedance spectroscopy and (ii) electrochemical impedance spectroscopy. The solid-state impedance spectroscopy was first utilized in high-temperature solid oxide fuel cells to observe its electrical properties as a function of frequency.[11] The method has been applied to a wide range of applications such as detecting corrosion layers on metallic materials, determining the ionic conductivity for different components in batteries, fuel cells, and solar cells, dielectric polarization in capacitors, conductive percolation in composites, etc. In this system, dielectric materials were sandwiched between two metal contacts and various aforementioned electrical phenomena originated from various contributions such as grain, grain boundary, and electrode ceramic interface effect. Later, electrochemical impedance spectroscopy became the most widely used technique which uses a liquid electrolyte to assess how the materials being tested and reacted when exposed to the liquid electrolyte. This procedure is more commonly referred to as electrochemical impedance spectroscopy because it is an electrochemical procedure. This is the approach that helps with comprehending supercapacitor, battery, fuel cell, chemical, or bio-sensor behavior for a variety of uses. The concentration of electroactive species, charge, and mass transfers from the bulk solution to the electrode surface, and electrolyte resistance are the inferences of matter-(redox species)-electrode interactions in a standard electrochemical cell.[12] As seen in **Figure 2**, each of these characteristics is represented by an electrical circuit made up of resistors, capacitors, or constant phase elements coupled in series or parallel to create an equivalent circuit. The equivalent circuit consists of three main components- resistance (R), capacitance (C), and inductance (L). The opposition of current (I) flow through a circuit is known as a resistor by Ohm's law and it can be represented by,

$$R = \frac{V}{I} \ldots (7)$$

A capacitor does not have constant current flowing through it. However, the capacitance (capacitance C in farad, F) of the capacitor depends on the supplied constant voltage and stores an amount of electrical charge (q), which is equal to the product of the applied potential and its capacitance,

$$Q = CV \ldots (8)$$

If an ac voltage $V(t) = V_0\sin(\omega t)$ is applied through the capacitor, the current will be,

$$I = \frac{dQ}{dt} = V_0 C\omega\cos(\omega t) = I_0\cos(\omega t)$$

Where, $I_0 = V_0 C\omega$ and the capacitive reactance is, $X_c = \frac{V_0}{I_0} = \frac{1}{\omega C} = \frac{1}{2\pi f_c}$

The voltage applied across an inductance (L) is provided by the equation,

$$V = L\frac{di}{dt} \ldots (9)$$

In a similar way, we can find the inductance reactance $X_L = \omega L$

The phaser diagram of these three components of an equivalent circuit model is shown in **Figure 3**.

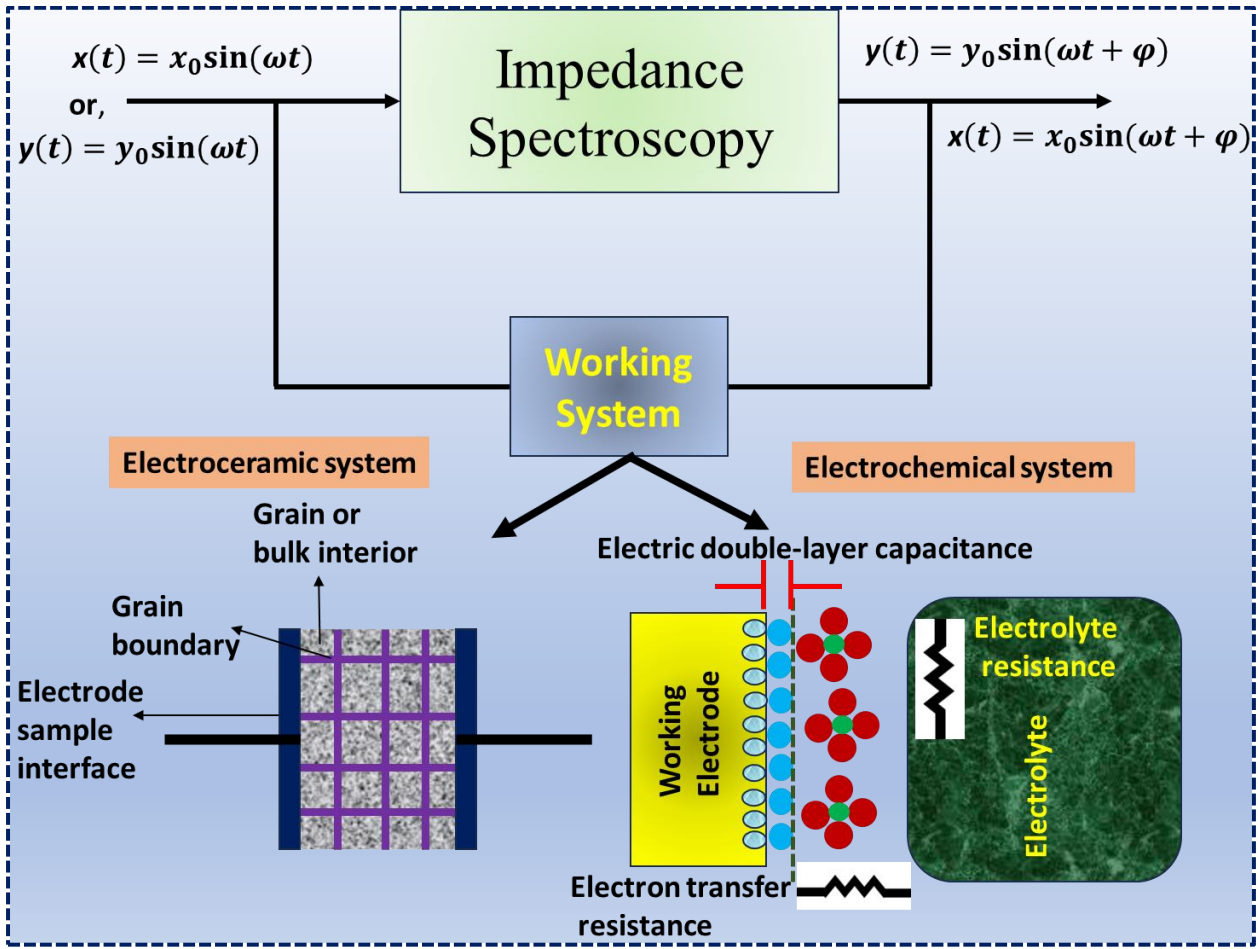

*Figure 2. Schematic illustration of impedance spectroscopy with its working system.*

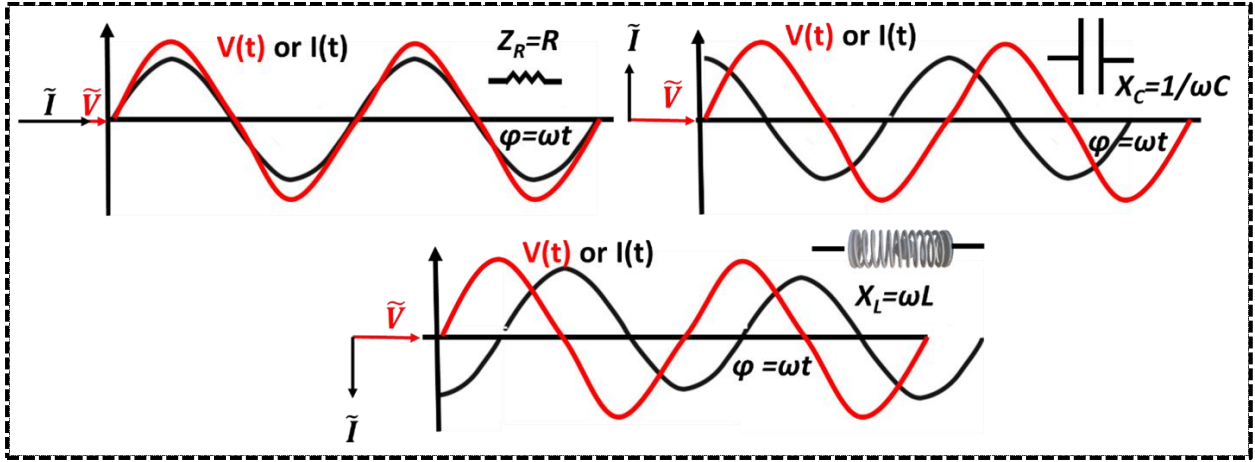

*Figure 3. The phaser diagram of resistor, capacitor, and inductor applying sinusoidal AC current or voltages.*

## 3. Impedance Spectroscopy for Electroceramic System

Solid-state impedance spectroscopy is used to characterize and analyze entire electronic, dielectric, semiconducting, and superconducting properties of all kinds of materials, such as bulk ceramics, polymers, semi-metals, thin-film, and composites. With the use of this adaptable tool, scientists and engineers can investigate the capacitance, conductance, and impedance of different material systems over a broad frequency as well as temperature range which is important for maximizing material performance and creating cutting-edge goods for a variety of markets, such as energy storage, electronics, and telecommunications. The solid-state impedance analyzer calculates various factors such as impedance, phase, capacitance, loss, admittance, conductance, susceptance, etc. by applying a small AC voltage to the materials (mostly dielectric or semiconducting/semi-metal) under test and measuring the resulting current. The frequency generator produces AC voltage at various frequencies and the output electrical parameters exhibit two contributions of the real and imaginary part.[13] The evolution of various impedance parameters with frequency as well temperature (above and below room temperature) of any materials system in the shape of parallel pellet capacitor or thin film is one of the widely researched areas and we will discuss here briefly those parameters with formulism and analysis.

### 3.1. Impedance Formulism

The real and imaginary impedance of any materials under test can be calculated by eq$^n$ (5) & (6) at various temperatures and it usually contains two different regions- (i) frequency independent region associated with the DC conduction, and (ii) frequency-dependent AC conduction region.

The long-range translational movement of charge carriers is responsible for DC conduction, and it is possible due to their successful and effective hopping which causes their neighboring charge particle to relax to their position.[14,15] The DC conduction is mostly visible in the real part of impedance spectra, and it is shown in **Figure 4**a by one typical plot. The charge carriers travel locally because of their unsuccessful hopping, which allows them to settle into their location and leads to ac conductivity in the impedance plot. These two different forms of material movement result in the appearance of DC resistivity in the lower frequency zone and frequency-dependent AC conductivity in the dispersive region due to the conductive nature of the substrate. It has been observed that at a specific frequency, the Z' value loses its frequency dependence and shifts towards the higher frequency zone as temperatures rise. This behavior suggests that the produced material may be undergoing a process of frequency relaxation process. The imaginary impedance in **Figure 4**b is frequently associated with dielectric relaxation processes in dielectric materials which are associated with the migrations of charge carriers or orientation of the polar molecules by the application of an externally applied electric field. In some materials, the imaginary impedance is entangled with frequency-dependent conductivity, which exhibits variable-range hopping or other hopping conduction. The applied AC signal frequency determines how quickly charge carriers transition between localized states. Also, the active ions moving in response to an applied electric field can aid in relaxation processes in materials possessing ionic conductivity and produce a relaxation plateau in the imaginary impedance spectra which is associated with the time taken by the ions to migrate and rearrange.[16,17]

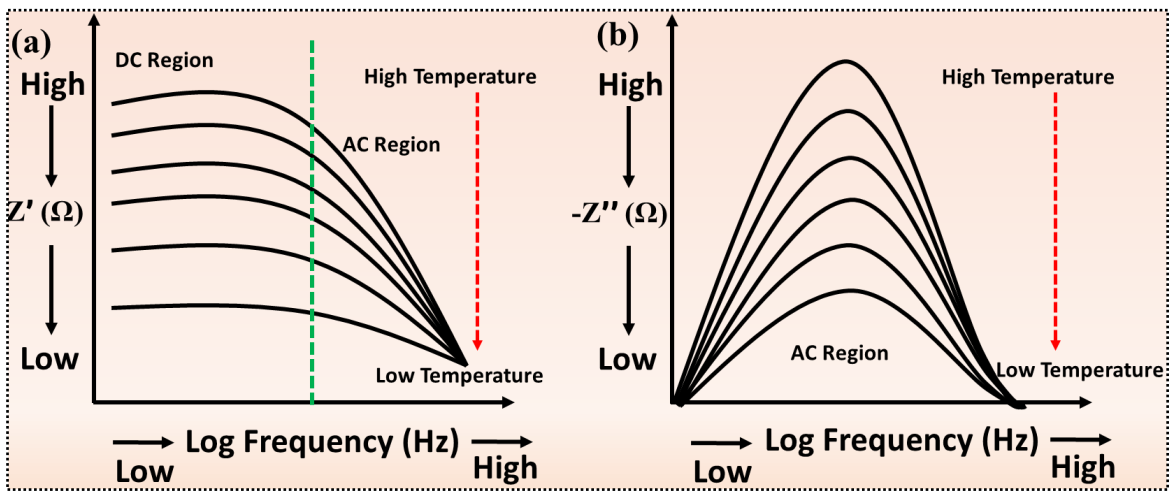

*Figure 4*. One typical plot of ceramic materials is (a) real impedance with frequency, and (b) imaginary impedance with frequency at different temperatures.

In polycrystalline materials, the electrical charge transport mechanism and the macroscopic dielectric constant are typically composed of three elements, grain or bulk contribution, grain boundary contribution, and the electrode-ceramic interface effect. To obtain an accurate comprehension of the operation of a polycrystalline electroceramic, it is imperative to dissect these contributions utilizing a suitable comparable equivalent network circuit model to characterize and evaluate every relaxation independently. This model is known as the bricklayer model which summarizes the various presumptions that must be made to use impedance spectroscopy in this manner. It is not necessary that all grain boundary areas across the sample show similar properties, such that they can be averaged by a single equivalent circuit component, and the grain boundaries show distinctively different properties compared to grain interior regions or other contributions to allow discrimination to account for potentially different dielectric and resistive behavior in grain boundaries compared to bulk regions.[18,19] Naturally, the same reasoning applies to all other contributions, including grain inside and interface. These factors make it abundantly evident that impedance spectroscopy analysis on non-uniform materials must be done very carefully. It is also suggested that if the inner grain contributions are noticeably more resistive than grain boundary regions, they do not show up in the impedance spectra. Only the conducting grain boundary contribution can be analyzed in terms of resistive and dielectric behavior when very resistive grains are excluded which can be observed in single crystalline wide band gap semiconductors. However in meticulously manufactured homogeneous polycrystalline electroceramic materials, the grain boundary regions are frequently more resistive than the grains, and the impedance spectra should exhibit all three of the contributions.[20] According to the prepared samples, there is a possibility to observe each contribution or their combined network mode. So, we will discuss the most popular network models used in polycrystalline or single crystalline samples by Nyquist ($Z'$ vs. $Z''$) or Bode plots, and each contribution mostly contains an RC network model (R-resistance, C-capacitance). Sometimes, a constant phase element (CPE) is used in place of a resistor or capacitor, and it is associated with a phase element and coefficient which decides its behavior as resistive or capacitive. For single contributions like grain or grain boundary, only one equivalent circuit containing one resistor and capacitor in parallel is used and the overall impedance formula can be written as[21,22]

$$Z(\omega) = \frac{1}{\frac{1}{R_1}+j\omega C_1} = \frac{R_1}{1+j\omega C_1 R_1} = \frac{R_1}{1+(\omega C_1 R_1)^2} - j\frac{\omega R_1^2 C_1}{1+(\omega C_1 R_1)^2} \quad \dots (10)$$

and the Nyquist plots corresponding to the semicircles are shown in **Figure 5**a. At higher frequencies, $\omega$ tends to $\infty$ and $Z(\omega)$ tends to zero and current passes through the capacitor. At lower frequencies, $\omega$ tends to 0 and $Z(\omega)$ tends to $\infty$, and current passes through the resistor. There are only relaxation peaks observed in the imaginary impedance spectra corresponding to the maximum impedance ($\omega Z''_{max}$) and the characteristics time constant was obtained $\tau_1=R_1C_1$. If the RC network contains another resistance $R_0$ in series, the resulting equation for the circuit will be,[23]

$$Z(\omega) = R_0 + \frac{R_1}{1+(\omega C_1 R_1)^2} - j\frac{\omega R_1^2 C_1}{1+(\omega C_1 R_1)^2} \quad \dots (11)$$

and the semicircle in the Nyquist plot starts after an interval along the X-axis, shown in **Figure 5**b. The additional resistance originates due to contact between the material surface and the electrode terminal.

When the materials or ceramics have both grain and grain boundary contributions, two semicircles appeared in the Nyquist plots with different contributions of R and C. The circuit contains two RC networks connected in series with additional contact resistance and the equivalent circuit with Nyquist plot and impedance shown in **Figure 5**c. The impedance of the equivalent circuit will be,

$$Z(\omega) = [R_0 + \frac{R_1}{1+(\omega C_1 R_1)^2} + \frac{R_2}{1+(\omega C_2 R_2)^2}] - j[\frac{\omega R_1^2 C_1}{1+(\omega C_1 R_1)^2} + \frac{\omega R_2^2 C_2}{1+(\omega C_2 R_2)^2}] \quad \dots (12)$$

The circuit also contains two different time constants $\tau_1=R_1C_1$ and $\tau_2=R_2C_2$ corresponding to grain and grain boundary, respectively. When $\tau_1=\tau_2$, only one contribution either grain or grain boundary exists in the materials. According to the general rule, the peak that emerged at the higher frequency side at room temperature corresponds to the grain, and the peak that formed at the lower frequency side corresponds to the grain boundary effect. The $Z''(\omega)$ spectrum and phase both show two peaks in **Figure 5**d. Similarly, three semicircles in the Nyquist plots indicate the presence of grain, grain boundary, and electrode ceramic interface effect. The higher frequency semicircle is attributed to the grain effect, the intermediate semicircle is attributed to the grain boundary effect, and the lower frequency semicircle is attributed to the electrode ceramic interface effect. The equivalent circuit for this kind of material will be,

$$Z(\omega) = [R_0 + \frac{R_1}{1+(\omega C_1 R_1)^2} + \frac{R_2}{1+(\omega C_2 R_2)^2} + \frac{R_3}{1+(\omega C_3 R_3)^2}] - j[\frac{\omega R_1^2 C_1}{1+(\omega C_1 R_1)^2} + \frac{\omega R_2^2 C_2}{1+(\omega C_2 R_2)^2} + \frac{\omega R_3^2 C_3}{1+(\omega C_3 R_3)^2}]$$
$$\dots (13)$$

The radii of each semicircle are associated with the unique relaxation process and the time constant corresponds to the three contributions. The Debye-type conduction process is indicated by a perfect semicircle whose center lies on the abscissa. If the semicircle suppresses and its center does not lie on the abscissa, the non-Debye type of conduction comes into play. In the non-Debye type of conduction, all diploes within the dielectric materials are not oriented in a particular direction at a time.[24,25] Usually, the radius of semicircles decreases with high-temperature evolution which signifies the increase in conductivity in the materials.

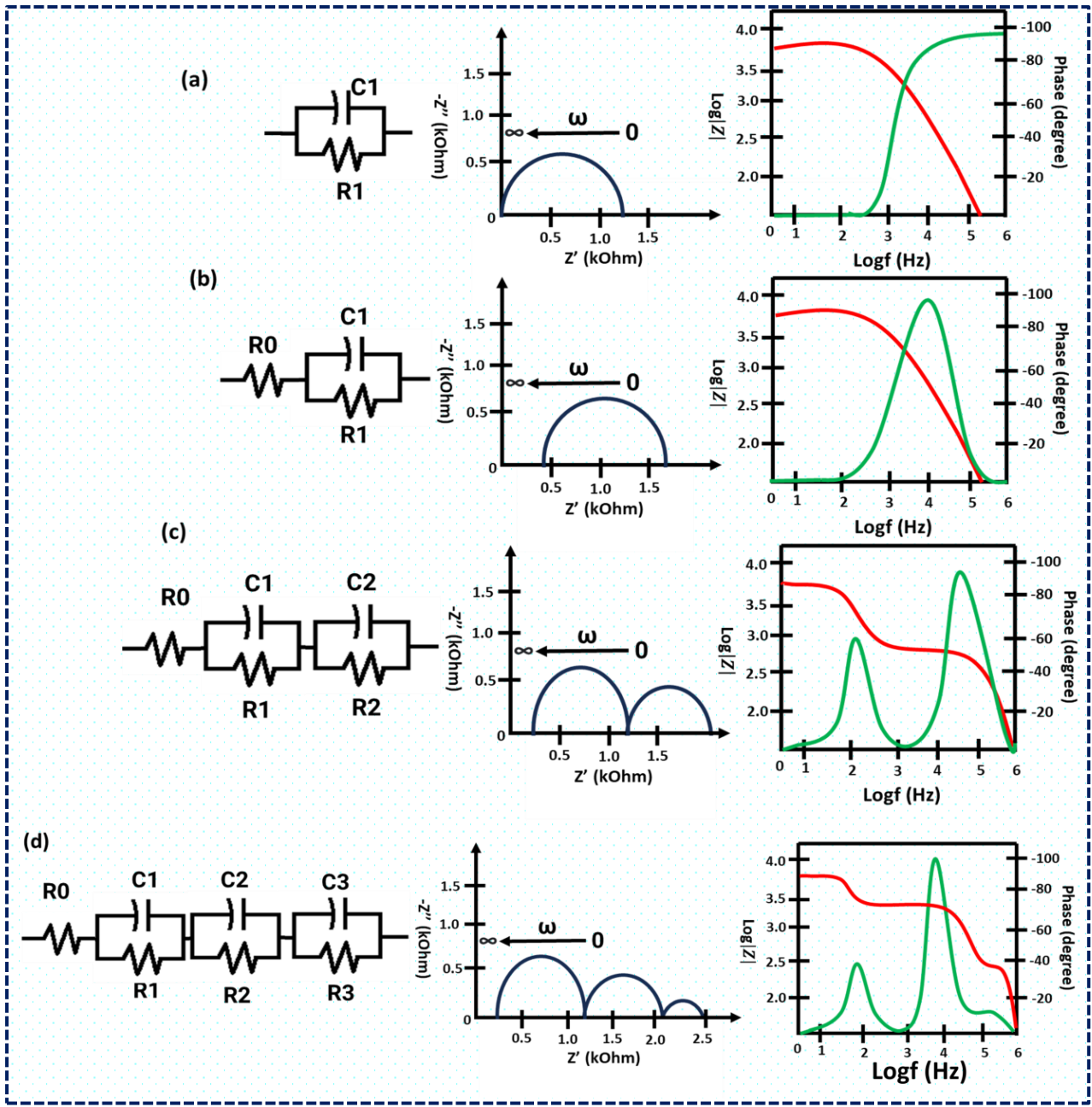

*Figure 5. The equivalent network, Nyquist plots, and Bode impedance and phase angle of some model circuits (a) $R_1C_1$ for grain contributions, (b) $R_0(R_1C_1)$ for grain contributions with contact resistance, (c) $R_0(R_1C_1)(R_2C_2)$ for grain and grain boundary contributions with contact resistance, (d) $R_0(R_1C_1)(R_2C_2)(R_3C_3)$ for grain, grain boundary and electrode ceramic interface effect with contact resistance.*

The fractional equivalent circuit models are frequently used to simulate the behavior of capacitors, batteries, and supercapacitors in the frequency and time domains due to more accuracy by using constant phase elements (CPEs). Most of the experimental data acquired with electroceramic or electrochemical impedance spectroscopy are included in these analogous circuits.[26,27] The CPEs are constant, and they depend on a fractional coefficient (α) by,[28]

$$Z_{CPE}(j\omega) = \frac{1}{Q.(j\omega)^\alpha} = \frac{1}{Q\omega^\alpha} \exp\left(-j\alpha \frac{\pi}{2}\right) \ldots (14)$$

The shape of the semicircle in the Nyquist plots depends on the values of the fractional coefficient and it is usually less suppressed than a regular RC network. For a single contribution of grain or grain boundary, the complex impedance of a circuit (**Figure 6**a) can be written as,

$$Z^*(\omega) = \frac{R_1}{1+R_1Q_1(j\omega)^{\alpha_1}} \ldots (15)$$

For an additional contact resistance (**Figure 6**b) $Z^*(\omega) = R_0 + \frac{R_1}{1+R_1Q_1(j\omega)^{\alpha_1}} \ldots (16)$

For grain and grain boundary contribution (**Figure 6**c), $Z^*(\omega) = R_0 + \frac{R_1}{1+R_1Q_1(j\omega)^{\alpha_1}} + \frac{R_2}{1+R_2Q_2(j\omega)^{\alpha_2}}$
… (17)

For grain, grain boundary, and ceramic interface effect (**Figure 6**d),

$$Z^*(\omega) = R_0 + \frac{R_1}{1+R_1Q_1(j\omega)^{\alpha_1}} + \frac{R_2}{1+R_2Q_2(j\omega)^{\alpha_2}} + \frac{R_3}{1+R_3Q_3(j\omega)^{\alpha_3}} \ldots (18)$$

The coefficients $\alpha_1$, $\alpha_2$, and $\alpha_3$ correspond to the grain, grain boundary, and electrode ceramic interface effect, respectively. The capacitance of all three contributions can be calculated from the fractional coefficient by a generalized relation,

$$C = R^{\left(\frac{1}{\alpha}-1\right)} Q^{\frac{1}{\alpha}} \ldots (19)$$

An essential tool for modeling and analyzing impedance data is the Constant Phase Element (CPE). It offers a more realistic depiction of distributed processes and non-ideal capacitive behavior in a variety of systems, including semiconductor devices and electrochemical interfaces. Researchers and engineers can obtain greater insights into the characteristics and functionality of materials and devices, resulting in improved design and optimization techniques, by adding the CPE into similar circuit models.

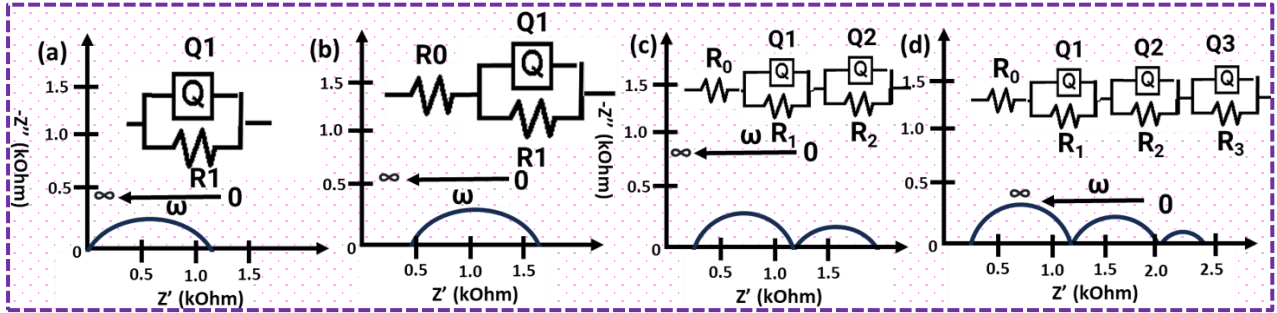

**Figure 6**. *The Nyquist plots and equivalent network model of (a) $R_1Q_1$ for grain contributions, (b) $R_0(Q_1R_1)$ for grain contributions with contact resistance, (c) $R_0(Q_1R_1)(Q_2R_2)$ for grain and grain boundary contributions with contact resistance, (d) $R_0(Q_1R_1)(Q_2R_2)(Q_3R_3)$ for grain, grain boundary and electrode ceramic interface effect with contact resistance.*

**3.2. Modulus Formalism**

The modulus formalism, sometimes referred to as the magnitude spectrum in impedance spectroscopy, offers important details regarding the amplitude or magnitude of the impedance response at various frequencies as well as temperatures. Understanding the electrical characteristics of materials and devices requires an understanding of the modulus spectra. Impedance spectroscopy relies highly on modulus spectra for the following reasons:[29,30]

(i) The modulus spectrum aids in distinguishing the contributions of reactive and resistive elements in complex impedance spectra. The modulus, which can be used to indicate the magnitude of impedance at a particular frequency, can be used to determine the relative importance of resistive, capacitive, and inductive elements by breaking it down into real and imaginary components. In the imaginary impedance loss spectrum, the grain or grain boundary or electrode ceramic interface effects are less pronounced compared to the other two and it can be a challenging part to distinguish. The reduced or suppressed capacitance/impedance value is highlighted by electric modulus spectra, which will amplify the grain or grain boundary or electrode ceramic effects and make them easier to examine.

(ii) The major electrical processes present in a material can be determined with the use of the modulus spectra. In the magnitude spectrum, certain relaxation processes, conductivity mechanisms, or impedance contributions are represented by peaks or distinctive features. Understanding the underlying physics of the material is aided by analyzing these traits.

(iii) Optimizing material qualities for use in electronic devices or sensors requires an understanding of modulus spectra. Based on the obtained modulus spectra, researchers can adjust material compositions and manufacturing conditions to achieve desired electrical characteristics.

(iv) The modulus spectra are essential for analyzing electrochemical impedance data in electrochemical systems, where impedance spectroscopy is frequently employed. They facilitate knowledge of ion transport phenomena, charge transfer procedures, and the kinetics of electrochemical reactions.

The complex electrical modulus ($M^*$) is represented by the reciprocal of complex permittivity ($\varepsilon^*$) as well they have a direct relationship with complex impedance spectra by,[31,32]

$$M^* = \frac{1}{\varepsilon^*} = M' + iM'' \dots (20)$$

$$M' = \omega C_0 Z'' = \frac{\varepsilon'}{\varepsilon'^2 + \varepsilon''^2} \dots (21)$$

$$M'' = \omega C_0 Z' = \frac{\varepsilon''}{\varepsilon'^2 + \varepsilon''^2} \dots (22)$$

where, $\varepsilon'$ and $\varepsilon''$ are the real and imaginary parts of dielectric permittivity, respectively. The terms $C_0 = \frac{\varepsilon_0 A}{d}$, $\varepsilon_0 = 8.854 \times 10^{-12}\ F/m$, A, and d is the area and thickness of the samples. From one typical plot of $Z''$ and $M''$ with frequency, shown in **Figure 7**a, it is observed that the suppressed region of $Z''$ in the high-frequency regions is highlighted in the $M''$ spectra with temperature evolution. To distinguish the contribution between the grain, grain boundary, and electrode ceramic interface effect, the Cole-Cole plot of modulus ($M'\ vs.\ M''$) is illustrated in **Figure 7**b. In contrast to the two semicircles observed in Cole–Cole of impedance, Cole–Cole of modulus displays a single semicircle with a small tail for the whole temperature range under investigation. Sometimes, the suppressed semicircle contains two contributions such as grain boundary and grain with a small contribution from the electrode interface effect, associated with the small tail. When comparing the Cole-Cole plot of modulus and impedance at room temperature, we can see that the feature of a complete semicircle joined by an arc segment appears in both scenarios. The presence of an arc segment, however, which relates to a high-frequency zone in the Cole-Cole of modulus and a low-frequency zone in the Cole-Cole of impedance, is what makes the difference.

If the imaginary modulus ($M''$) vs. frequency exhibits one relaxation peak, as shown in **Figure 7c**, which will correspond to either grain or grain boundary. In this case, the measured peaks are wider than the Debye peaks and exhibit a small asymmetry at each temperature. The expansion with temperature confirms the non-Debye relaxation of any typical dielectric materials by highlighting the distribution of relaxation with various meantime constants. Furthermore, the temperature-induced shift in peak positions indicates a move toward higher frequencies. Usually, the single or double relaxation peaks in the modulus spectra can be demonstrated by the numerical Laplace transform of the Kohlrausch–Williams–Watts (KWW) decay function. The frequency-dependence complex electric modulus is related to the time-dependent electric field $\varphi(t)$,[33,34]

$$\varphi(t) = \exp\left[-\left(\frac{t}{\tau_{max}}\right)^{\beta}\right]; (0<\beta'<1) \ldots (23)$$

and,

$$M^*(\omega) = M_\infty\left[1 - \int_0^\infty \exp(-i\omega t)\left(-\frac{d\varphi(t)}{dt}\right)\right]dt \ldots (24)$$

The parameter $\beta'$ is known as the Kohlrausch parameter ($0<\beta'<1$) and it is related to the full width at half maximum (FWHM) of the relaxation curves; $\tau_{max}$ is the relaxation corresponding to the $\omega_{max}$. The curves in **Figure 7c** exhibit one single peak characteristic and it can be fitted by the well-known Bergman equation,

$$M'' = \frac{M''_{max}}{[(1-\beta')+(\frac{\beta'}{1+\beta'})\{\beta'(\frac{\omega_{max}}{\omega})+(\frac{\omega}{\omega_{max}})^{\beta'}\}]} \ldots (23)$$

Where $M''_{max}$ is the maximum modulus associated with the maximum frequency $\omega_{max}$. In **Figure 7d**, two different relaxation peaks are observed in the imaginary modulus spectra which correspond to the grain and grain boundary contributions with two distinct regions of different slopes. The imaginary modulus plots can be fitted with,[35]

$$M'' = \frac{M''_{1max}}{[(1-\beta_1')+(\frac{\beta_1'}{1+\beta_1'})\{\beta_1'(\frac{\omega_{1max}}{\omega})+(\frac{\omega}{\omega_{1max}})^{\beta_1'}\}]} + \frac{M''_{2max}}{[(1-\beta_2')+(\frac{\beta_2'}{1+\beta_2'})\{\beta_2'(\frac{\omega_{2max}}{\omega})+(\frac{\omega}{\omega_{2max}})^{\beta_2'}\}]} \ldots (24)$$

if the two Kohlrausch parameters $\beta_1'$ and $\beta_2'$ are mutually independent. All relaxation phenomena are caused by thermal energy, and these dynamical growths are strongly temperature-dependent. The maximum angular frequency ($\omega_{max}$) is reciprocal to the relaxation time ($\tau_{max}$) and activation energy ($E_m$) can be calculated by[36,37]

$$\omega_m = \frac{1}{\tau_{max}} = \omega_0 \exp\left(-\frac{E_m}{k_B T}\right) \ldots (25)$$

Where $E_m$ is the activation energy corresponding to grain, grain boundary, or electrode interface, and $k_B \sim 8.6173 \times 10^{-5}$ eV K$^{-1}$ is the Boltzmann constant.

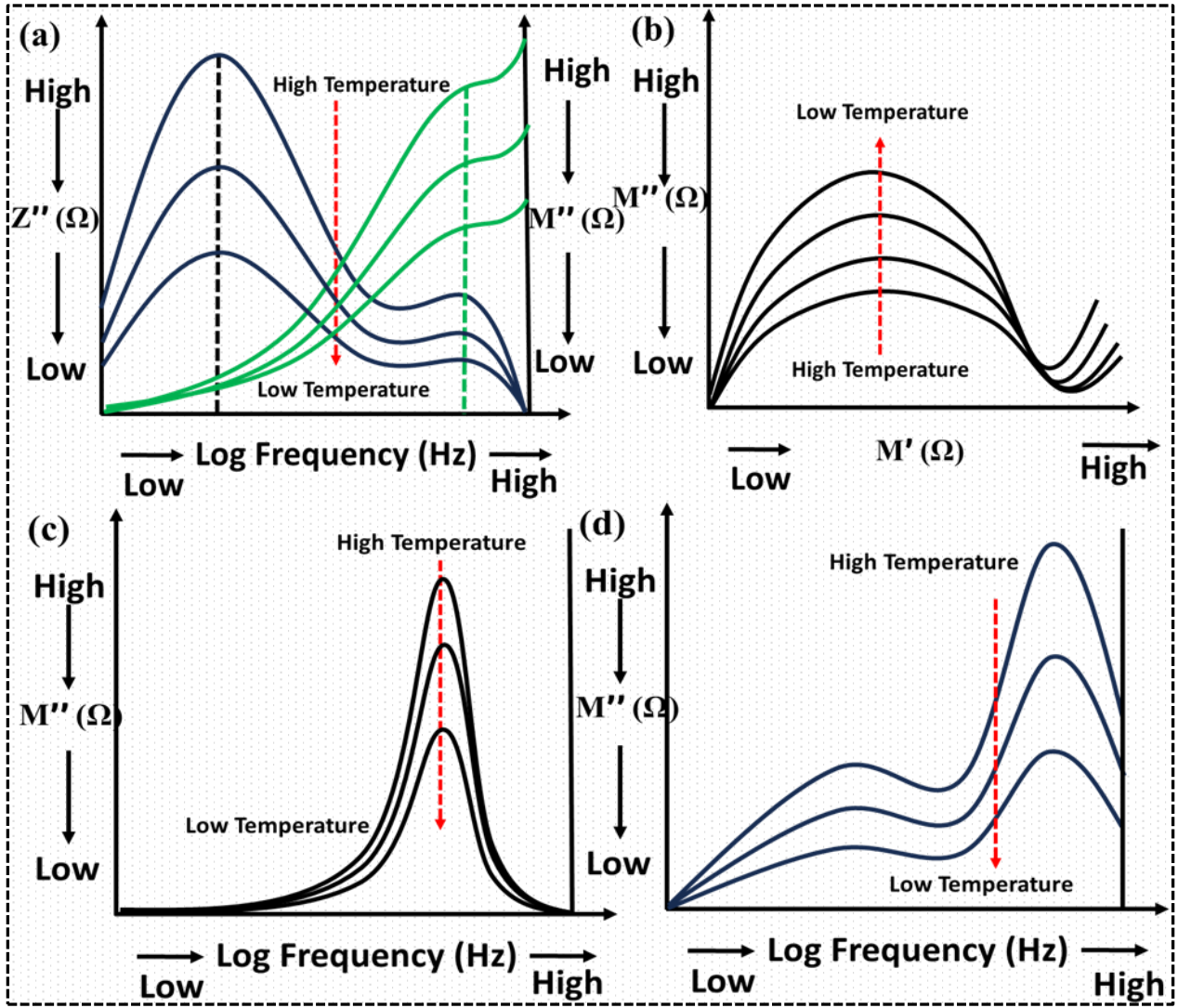

*Figure 7. Typical plots of (a) $Z''$ and $M''$ with frequency, (b) Cole-Cole plot ($M'$ vs. $M''$) of modulus at different high temperatures. (c) single relaxation peaks, (d) double relaxation peaks in the $M''$ vs. frequency.*

### 3.3. Dielectric Formulism

The electrical behavior of any material in the presence of an external electric field is described by two crucial parameters: the dielectric loss and the dielectric constant, which are also referred to as the relative permittivity. In the domains of electrical engineering, materials science, and electronics, these electrical qualities are very important. An outline of the significance of dielectric constant and dielectric loss is provided below: -

The capacitance of any substance in the electronic components, like capacitors, is determined by its dielectric constant. Increased capacitance results from greater dielectric constants, which is advantageous in situations requiring a larger capacity for charge or energy storage. High relative permittivity dielectric materials are employed in energy storage applications because they improve the capacity to store electrical energy which is important for technologies like pulse power systems and energy storage capacitors. When designing and operating dielectric resonators and antennas, the dielectric constant is crucial, and it affects the radiation efficiency, impedance matching, and resonant frequency of the components in the communication systems. The dielectric materials having a high dielectric constant lessen the intensity of the electric field inside the material when an external voltage is applied. In order to prevent electrical breakdown, it is beneficial for dielectric materials used in the electrical insulation. Dielectric loss in capacitors causes heat and energy losses, which can lower the overall efficiency of an electronic circuit. In power systems, dielectric loss adds to the power factor, and the efficiency with which electrical power is transformed into productive work is measured by the power factor. To reduce the energy losses in the power distribution systems, dielectric loss control is essential. The quality factor (Q factor) of any resonant circuit is greatly impacted by dielectric loss because the energy storage and dissipation of the circuits are controlled and measured using the Q factor.[38–40]

The dielectric material (sometimes known as an electrical insulator) becomes polarized when an external electric field is applied to it. This causes the positive charges to move toward the negative electrode and the negative charges to orient themselves toward the positive electrode. The polarization effect that resists the applied field pushes charges onto the electrodes since charges are not free to flow in an insulator. The quantity of charge that can be stored in the capacitor increases with the ease of polarization of the material. The dielectric constant, also known as relative permittivity ($K$ or $\varepsilon_r$), is the measure of the energy storage capacity of any material under an electric field. Some external factors such as mechanical stress or the action of an electric field are responsible for producing electrical polarization in the piezoelectric crystals which are in solid forms that can store electric charge. Dielectric Polarization can also happen on its own in pyroelectric crystals, particularly in ferroelectrics which allows an electric field to be applied to reverse the spontaneous electric polarization. The degree of polarization (P) is strongly dependent on the dielectric constant by,[41,42]

$$P = \varepsilon_0(\varepsilon_r - 1)E \ldots (26)$$

The parameters $\varepsilon_0$ and E represent the vacuum space permittivity and electric field strength. Mostly, four types of electrical polarizations i.e., electronic, ionic, dipolar or orientational, and interfacial or space charge polarizations occur in any dielectric medium under an electric field and the sum of all polarization gives overall polarizations on the system. **Figure 8**a represents different polarization with frequency from a typical real, imaginary dielectric permittivity and tangent loss spectra.

The electronic polarizations have no impact on temperature and as temperatures rise, the dielectric constant decreases. The dielectric thermal expansion is the reason behind it and when an electric field is applied, the nucleus of atoms and the center of the electron clouds move apart, forming a tiny dipole with a very small polarization impact. This kind of polarization can be seen in the very high-frequency zone above $10^{14}$ Hz, and it can be measured by broadband dielectric spectroscopy. The thermal factor of the dielectric constant is, $\alpha_T = \frac{1}{\varepsilon_r}\frac{d\varepsilon_r}{dt}$. In the case of ionic polarizations, ions are symmetrically distributed in a crystal lattice with net zero polarization in ionic solids, such as ceramic ceramics. The cations and anions are drawn in opposing directions as soon as an electric field is applied. Compared to electronic displacement, this results in a comparatively large ionic displacement, which in turn leads to high dielectric constants in ceramics that are frequently used in capacitors. The electric dipole moments of two oppositely charge-shifted ions in ionic polarizations are shown in **Figure 8**b. In some substances, there are permanent molecular dipoles that rotate in the direction of the applied field when exposed to an electric field. This results in a net average dipole moment per molecule, and it is known as dipolar or orientational polarization. Polymers are more likely to exhibit dipolar orientation because of the ability to reorient the atomic structure. Interfacial or space charge polarizations occur in ceramics when impurities or irregular geometry at the interfaces of polycrystalline solids cause superfluous charges. This extrinsic form of polarization is brought about by these partially mobile charges that migrate under an applied field. All different electric polarizations occur at different frequency ranges as shown in **Figure 8**a.[43]

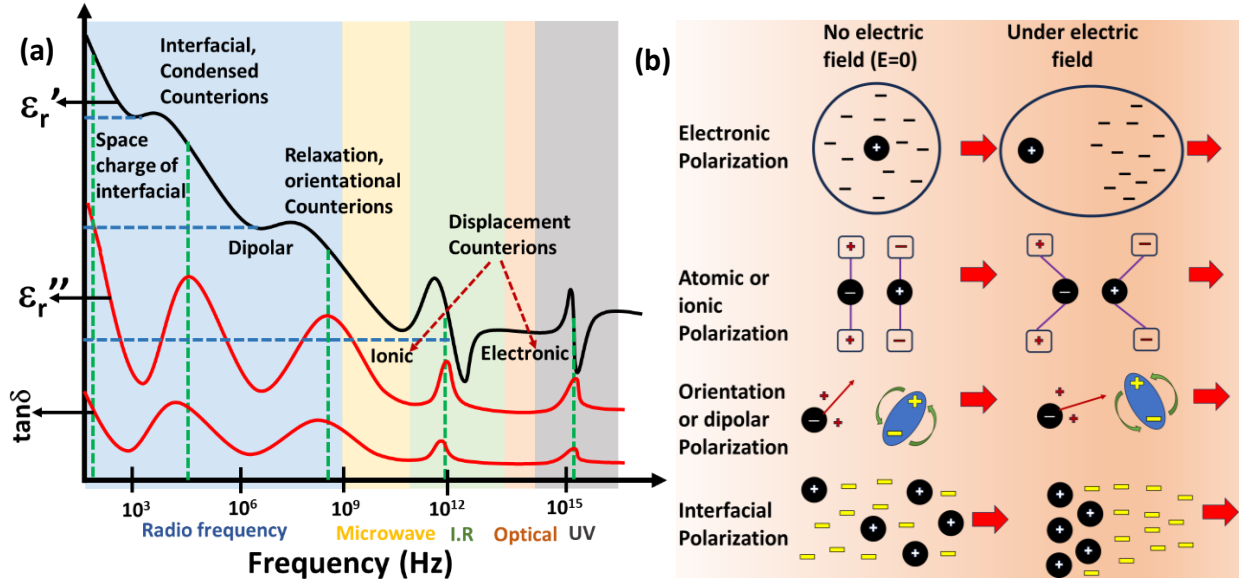

***Figure 8****. (a) The variation of the real part of dielectric ($\varepsilon'_r$), the imaginary part of impedance ($\varepsilon''_r$), and dielectric loss (tanδ) with frequency exhibits various electrical polarizations at various frequency zones. (b) Schematic diagram of the mechanism of electronic, atomic/ionic, orientational/dipolar, and interfacial polarization.*

Any important information about electrical polarization behavior, conduction mechanisms, and dielectric relaxation of any materials can be obtained by studying their dielectric properties. The permeability and permittivity of any materials are not constant, and they can vary depending on the molecular structure, combination, temperature, orientation, and pressure. The complex dielectric constant ($\varepsilon^*_r = \varepsilon'_r + i\varepsilon''_r$) has two components- real dielectric constant ($\varepsilon'_r$) and imaginary dielectric constant ($\varepsilon''_r$) and they can be calculated by two different formulas,[32,44,45]

$$\varepsilon'_r = \frac{C_p}{C_0} = \frac{C_p d}{A\varepsilon_0} = \frac{d}{\omega A \varepsilon_0} \frac{Z''}{Z'^2 + Z''^2} \quad \ldots (27)$$

$$\varepsilon''_r = \frac{d}{\omega A \varepsilon_0} \frac{Z'}{Z'^2 + Z''^2} \quad \ldots (28)$$

The $\varepsilon''_r$ has a direct relationship with $\varepsilon'_r$ by dielectric loss (tanδ),

$$\tan\delta = \frac{\varepsilon''_r}{\varepsilon'_r} \quad \ldots (29)$$

The dielectric constant with frequency decreases following different trends due to the inability of electric dipole moments to comply or adapt with the changes of applied AC electric field. The reorientation or realignment of electric dipoles within the material in response to variations in the electric field is known as dielectric relaxation. This phenomenon is especially significant for materials that contain ions or polar compounds. The behavior of dielectric materials in response to an applied electric field is described by various dielectric relaxation models, specifically in relation to the time-dependent reorientation of electric dipoles. The dielectric relaxation of different materials has been explained by a variety of theoretical models. These are the following dielectric relaxation models which are frequently used in oxides or electroceramics: -

(i) *Debye relaxation model*: A popular theoretical model for explaining how polar materials behave when an electric field is applied to them is called the Debye relaxation model. It was created by Peter Debye and is frequently used to comprehend how electric dipoles in dielectric materials reorient with time. The Debye model works best with straightforward and well-behaved dielectric systems. The complex electric permittivity ($\varepsilon^*_{r-Debye}$) in this model can be explained by,[46,47]

$$\varepsilon^*_{r-Debye} = \varepsilon_\infty + \frac{\varepsilon_0 - \varepsilon_\infty}{1 + i\omega\tau} \quad \ldots (30)$$

The real and imaginary part of dielectric permittivity can be expressed by,

$$\varepsilon'_{r-Debye} = \varepsilon_\infty + \frac{\varepsilon_0 - \varepsilon_\infty}{1 + \omega^2\tau^2} \quad \ldots (31)$$

and,
$$\varepsilon''_{r-Debye} = \frac{(\varepsilon_0 - \varepsilon_\infty)\omega\tau}{1 + \omega^2\tau^2} \quad \ldots (32)$$

Sometimes, the denominator $(1 + i\omega\tau)$ of eq$^n$(30) can be represented by $(1 - i\omega\tau)$ due to different sign conventions depending on the complex electric field, i.e., $(1 - i\omega\tau)$ for $\exp(-i\omega\tau)$ for or $(1 - i\omega\tau)$ for $\exp(+i\omega\tau)$. One typical plot for the Debye-type relaxation model is shown in **Figure 9**. The dielectric loss (δ) will be,[48]

$$\delta = \frac{\varepsilon''_r}{\varepsilon'_r} = \frac{(\varepsilon_0 - \varepsilon_\infty)\omega\tau}{\varepsilon_0 + \varepsilon_\infty\omega^2\tau^2} \quad \ldots (33)$$

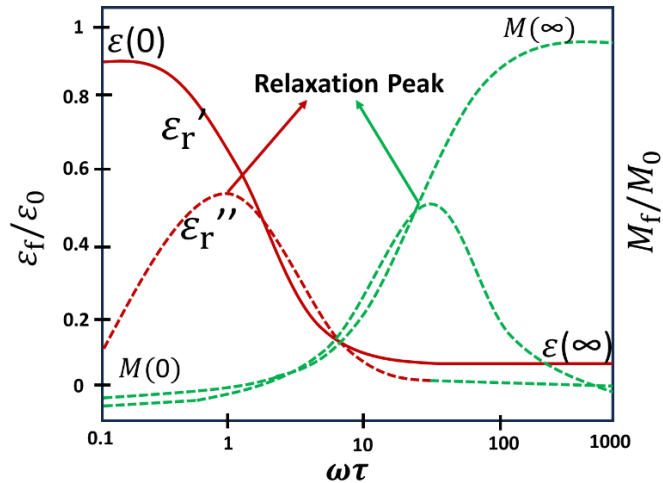

***Figure 9**. The Debye-type dielectric and modulus relaxation model in oxides.*

(ii) *Cole-Cole relaxation model:* The dielectric relaxation in materials with more complex characteristics, is described using the Cole-Cole model, which exhibits the distribution of relaxation time and is an extension of the Debye relaxation model. Robert H. Cole and Kenneth S. Cole created this model to explain the asymmetry in the dielectric spectrum. An extra parameter, α (known as the Cole-Cole parameter) is included in the model to describe the broadening of the dielectric response. According to this model, the complex electric permittivity ($\varepsilon^*_{r-(Cole-Cole)}$) can be represented as,[49,50]

$$\varepsilon^*_{r-(Cole-Cole)} = \varepsilon_\infty + \frac{\varepsilon_0 - \varepsilon_\infty}{(1+i\omega\tau)^{1-\alpha}} \quad \ldots (34)$$

The real and imaginary dielectric constant is obtained by solving the eq$^n$(34),

$$\varepsilon'_{r-(Cole-Cole)} = \varepsilon_\infty + (\varepsilon_0 - \varepsilon_\infty) \frac{1+(\omega\tau)^{1-\alpha}\sin(\frac{\alpha\pi}{2})}{1+2(\omega\tau)^{1-\alpha}\sin(\frac{\alpha\pi}{2})+(\omega\tau)^{2(1-\alpha)}} \quad \ldots (35)$$

$$\varepsilon''_{r-(Cole-Cole)} = \frac{(\varepsilon_0 - \varepsilon_\infty)(\omega\tau)^{1-\alpha}\cos(\frac{\alpha\pi}{2})}{1+2(\omega\tau)^{1-\alpha}\sin(\frac{\alpha\pi}{2})+(\omega\tau)^{2(1-\alpha)}} \quad \ldots (36)$$

The above two formulas can also be represented by introducing the hyperbolic function

$$\varepsilon'_{r-(Cole-Cole)} = \varepsilon_\infty + \frac{1}{2}(\varepsilon_0 - \varepsilon_\infty)[1 - \frac{\sinh(1-\alpha)\ln(\omega\tau)}{\cosh(1-\alpha)\ln(\omega\tau)+\sin(\frac{\alpha\pi}{2})}] \quad \ldots (37)$$

$$\varepsilon''_{r-(Cole-Cole)} = \frac{1}{2}(\varepsilon_0 - \varepsilon_\infty)[\frac{\cos(\frac{\alpha\pi}{2})}{\cosh(1-\alpha)\ln(\omega\tau)+\sin(\frac{\alpha\pi}{2})}] \quad \ldots (38)$$

Both dielectric constants in Cole-Cole relaxation can be reduced to the Debye model for α=0, and its value lies between 0 to 1. One plot containing real and dielectric constant following the Cole-Cole relaxation model is shown in *Figure 10*.

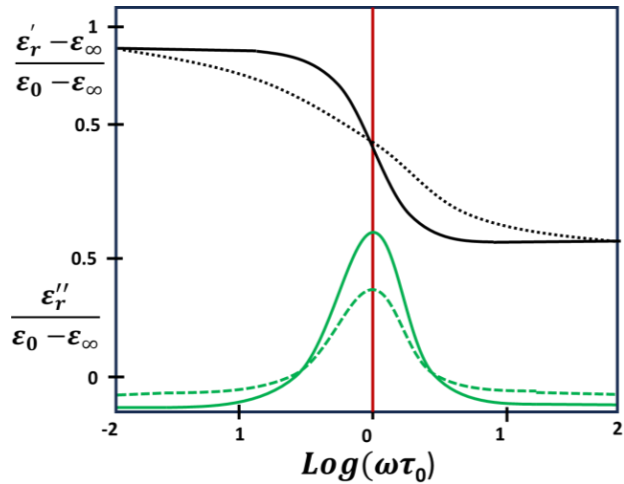

*Figure 10. The real and imaginary dielectric constant following the Cole-Cole relaxation model. The solid line represents the actual plots following this relaxation model and the dashed line frequently observed from experimental data.*

(iii) *Havriliak-Negami (HN) model:* A mathematical model called the Havriliak-Negami (HN) model is frequently used to explain dielectric relaxation in complicated materials including polymers, glasses, and biological tissues. In order to account for asymmetry and non-Debye behavior in the dielectric response, this model adds two parameters to the Debye and one parameter to the Cole-Cole models.

The complex dielectric permittivity in the HN model is given by,[51–53]

$$\varepsilon^*_{r-HN} = \varepsilon_\infty + \frac{\varepsilon_0 - \varepsilon_\infty}{[(1+i\omega\tau)^{1-\alpha}]^\beta} \quad \dots (39)$$

The additional parameter β is known as the broadening parameter and its value also lies between 0 to 1. According to this model, the real and imaginary can be calculated as,

$$\varepsilon'_{r-HN} = \varepsilon_\infty + (\varepsilon_0 - \varepsilon_\infty)[1 + 2(\omega\tau)^{1-\alpha}\cos\left(\frac{\alpha\pi}{2}\right) + (\omega\tau)^{2(1-\alpha)}]^{-\beta/2}\cos(\beta\varphi) \dots (40)$$

$$\varepsilon''_{r-HN} = (\varepsilon_0 - \varepsilon_\infty)[1 + 2(\omega\tau)^{1-\alpha}\cos\left(\frac{\alpha\pi}{2}\right) + (\omega\tau)^{2(1-\alpha)}]^{-\beta/2}\sin(\beta\varphi) \dots (41)$$

$$\text{where, } \varphi = \arctan\left(\frac{(\omega\tau)^{1-\alpha}\sin\left(\frac{(1-\alpha)\pi}{2}\right)}{1+(\omega\tau)^{1-\alpha}\cos\left(\frac{(1-\alpha)\pi}{2}\right)}\right)$$

In *Figure 11*, one can visualize the variation of real and imaginary dielectric constant following the Debye, Cole-Cole, and HN model**s**.

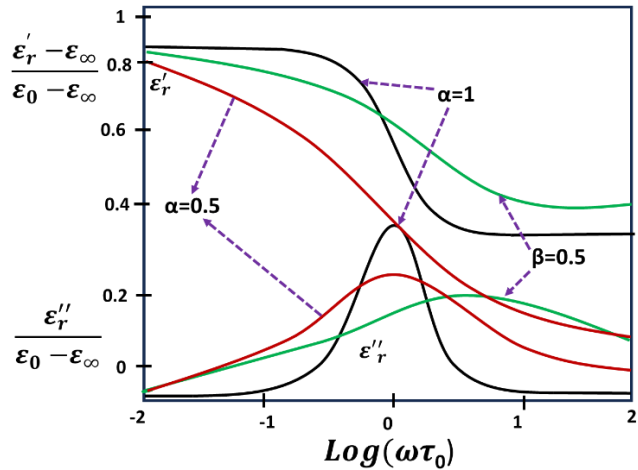

*Figure 11. Real and imaginary dielectric constant as a function of ωτ following the HN model with various parameters, α = 1 for Debye, α = 0.5 for Cole-Cole, β = 0.5 for HN relaxation.*

(iv) *Cole–Davidson model*: This model is a special case of the HN model for α = 0 and this relaxation model is usually used in the glass forming liquid materials.[54,55] The evolution of four fundamental dielectric relaxation with time is shown in *Figure 12.*

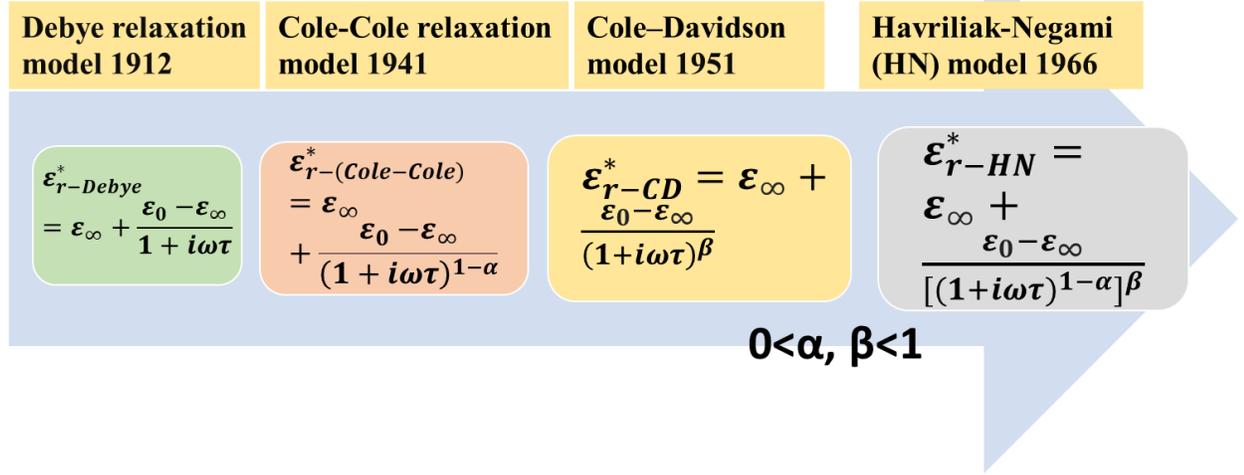

*Figure 12. The progression and development of different fundamental dielectric relaxation models.*

(v) *Kohlrausch-Williams-Watts (KWW):* Despite significant progress in recent years, a theoretical explanation of the slow relaxation in complicated condensed systems remains an active area of study. Slow relaxations appear to follow a universal function in the time domain. This function is called the Kohlrausch Williams-Watts (KWW) function, which is defined by a stretched exponential function,[56–58]

$$\varphi(t) = \exp\left[-\left(\frac{t}{\tau_{KWW}}\right)^\beta\right] \quad \dots (42)$$

where $\tau_{KWW}$ is the characteristic relaxation time and this function was 1$^{st}$ discovered in 1863 to explain the mechanical creep in glassy fibers. Later in 1970, Williams and Watts used this function to describe the dielectric relaxation phenomena in polymers. According to the relaxation model, the complex dielectric permittivity can be represented by,[59]

$$\varepsilon^*(\omega) = \varepsilon_\infty + (\varepsilon_0 - \varepsilon_\infty) \int_0^\infty \left[-\frac{d\varphi(t)}{dt} \exp(-i\omega t)\right] dt - i\frac{\sigma}{\omega} \quad \dots (43)$$

Using approximations, this eq$^n$(43) can be reduced to,

$$\varepsilon^*(\omega) \approx \varepsilon_\infty + \frac{(\varepsilon_0 - \varepsilon_\infty)}{\left[1 + \left(\frac{i\omega}{\omega_c}\right)^\alpha\right]^{\gamma(\alpha)}} - i\frac{\sigma}{\omega} \quad \ldots (44)$$

And the parameters, $\gamma(\alpha) = 1 - 0.812(1-\alpha)^{0.387}$; $\beta^{1.23} = \alpha\gamma(\alpha)$

$$\log_{10}(\tau) = -\log_{10}(\omega_c) - 2.6(1-\beta)^{0.5}\exp(-3\beta)$$

The conductivity (σ) term of eq$^n$(43) is associated with the conductivity-coupled dielectric relaxation process which can be added to the aforementioned relaxation model and correlated to the experimental data.

*Dielectric loss*

When an alternating electric field is applied to a dielectric medium, dielectric loss is used to describe the dissipation or loss of electrical energy and it is a measurement of an inefficiency of dielectric material in absorbing and releasing electrical energy. Dielectric relaxation processes in the material are also associated with dielectric loss. It can be defined as,[60,61]

$$\tan\delta = \frac{\varepsilon''}{\varepsilon'} \quad \ldots (45)$$

When the loss is indistinguishable due to free charge conduction σ,

$$\tan\delta = \frac{\omega\varepsilon'' + \sigma}{\varepsilon'} \quad \ldots (46)$$

An important consideration in the operation of electrical equipment like transformers, insulators, and capacitors is dielectric loss, and reducing dielectric loss is crucial to raising the efficiency of these devices. It guarantees that a higher percentage of electrical energy is efficiently transferred or stored without being dissipated as heat. The performance of electronic components i.e., printed circuit boards and integrated circuits components is also impacted by dielectric loss. Reducing dielectric losses is crucial in high-frequency applications to avoid signal deterioration and guarantee precise data transfer. The overall performance and efficiency of new technologies, such as energy storage systems (like supercapacitors), are also influenced by dielectric loss.

The electrode polarization in capacitive materials and Maxwell-Wagner-Sillars (MWS) polarization frequently dominate dielectric loss at low frequencies. The material that acts more like an ideal insulator may have a comparatively low dielectric loss. Different molecular and dipolar relaxation mechanisms become increasingly prominent in the middle-frequency range of dielectric loss spectra. Dielectric loss may result from these processes, and peaks or fluctuations in the loss tangent (tan δ) may correlate to particular relaxation frequencies. Ionic and electronic conduction might affect dielectric loss at high frequencies. Because there is less time for relaxation processes to take place during each cycle of the alternating field, the loss tangent may rise. Dielectric loss is thermally induced in many dielectric materials. Increased dielectric loss results from charge carriers being able to overcome energy barriers more easily at higher temperatures due to thermal energy. This kind of phenomenon is frequently seen in several ceramics and polymers. The pace and magnitude of dipolar and molecular relaxation processes are influenced by temperature. In the dielectric loss spectrum, higher temperatures can cause more noticeable relaxation peaks, particularly in materials with clearly defined relaxation processes. As the temperature varies, the dielectric loss peaks may shift which may be a sign of variations in the activation energy associated with particular relaxation mechanisms.

**3.4. Conduction Formulism:**

The AC conductivity spectra derived from impedance analysis are essential for comprehending how materials behave electrically across a spectrum of frequencies. The following salient features underscore the significance of AC conductivity spectra obtained from an impedance analyzer:[62–64]

(i) At certain frequencies, different materials show diverse conductivity responses, and comprehending this frequency dependence is crucial for describing the electrical characteristics of materials. Energy barriers related to charge carrier mobility are revealed by the temperature-dependent conductivity, which frequently follows an Arrhenius-type relation. Knowing this information is essential to comprehending how electrical conductivity in materials varies with temperature.

(ii) The dielectric characteristics of a substance are strongly related to its AC conductivity. Information regarding the polarization processes, including the contributions of ions and electrons, can be gleaned from the frequency-dependent behavior. This knowledge is essential for many applications, including the making of dielectric materials for capacitors. Temperature has a significant impact on conductivity, and it is possible to determine the characteristics of semiconductors, including bandgap energy, charge carrier concentration, and mobility, by examining ac conductivity spectra as a function of temperature.

(iii) Relaxation processes are frequently observed in the ac conductivity spectrum as peaks or shoulders, and they are related to the reaction of materials to an applied ac field which may be indicative of Maxwell-Wagner-Sillars effects, dipolar relaxation, or space charge polarization, etc. Certain temperatures can cause phase transitions in materials. By capturing variations in electrical behavior associated with the phase transitions, AC conductivity spectra can shed light on critical temperatures and the corresponding modifications to the structure and properties of materials.

(iv) The analysis of the contributions from ion migration is aided by the ac conductivity spectra for materials with significant ionic conductivity. It is essential to study the ac conductivity to understand electrolytes, ion-conductive polymers, and other materials used in fuel cells, batteries, and electrochemical devices. AC conductivity studies at various temperatures demonstrate the temperature-dependent ionic conductivity of solid electrolytes used in fuel cells and batteries. Optimizing the performance of these energy storage devices requires an understanding of temperature-dependent AC conductivity.

Therefore, the frequency and temperature-dependent AC conductivity is essential for improving the device performance, customizing materials for particular uses, and expanding our knowledge of charge transport pathways in a range of materials. The complex ac conductivity ($\sigma_{ac}^*$) is comprised of real ($\sigma_{ac}'$) and imaginary ($\sigma_{ac}''$) contributions and can be calculated by,[65–68]

$$\sigma_{ac}^* = \sigma_{ac}' + i\sigma_{ac}'' \ldots (47)$$

$$\text{where, } \sigma_{ac}' = \frac{t}{A}\frac{Z'}{Z'^2+Z''^2} \ldots (48)$$

$$\text{and, } \sigma_{ac}'' = \frac{t}{A}\frac{Z''}{Z'^2+Z''^2} \ldots (49)$$

The parameters t and A represent the thickness and area of the pallet capacitors, respectively. The AC conductivity is always associated with the frequency-dependent response of charge carriers of materials to an applied AC electric field. Charge carriers respond more dynamically at higher frequencies. The faster oscillation of the electric field at higher frequencies allows for more frequent interactions and contributions from different charge carriers and the AC conductivity frequently rises with frequency in materials where ionic conductivity is important, such as electrolytes or ionically conductive polymers. Increased conductivity results from ions having less time to travel across electrodes at higher frequencies. The ac conductivity in semiconductors or materials with localized charge carriers may be influenced by carrier hopping or tunneling between sites. Elevated frequencies could potentially lead to enhanced conductivity by enabling more effective hopping or tunneling procedures.

Under the influence of an AC electric field, weakly bound charged particles move in an orderly manner, resulting in thermally activated electrical conductivity in the materials. One of the important characteristics of the materials that need to be studied is how the charge carriers (i.e., electrons and holes, polarons, or cations and anions) respond to changes in temperature and frequency. These charge carriers control the majority of the conduction process, and the frequency dependence conductivity usually follows Jonscher's power law,[69,70]

$$\sigma_{tot}(\omega) = \sigma_{dc} + A\omega^n \dots (50)$$

The $\sigma_{dc}$ represents the frequency-independent dc conductivity, and $A$ and $n$ denote the unknown constant related to the strength of polarizability and frequency exponent, respectively. The long-range translational motion of ions or charge carriers contributes to DC conductivity which is responsible for the frequency-independent plateau at a low frequency region. In the jump relaxation model (JRM), Funke provided an explanation for the observed frequency-independent dc conductivity at low frequency and higher temperature and this hypothesis states that the conductivity at low frequencies is associated with the successful hops to nearby empty sites because of the extended time interval available; these subsequent jumps cause ions to translate over large distances, which contributes to dc conductivity. Two competing relaxation processes are usually involved at high frequencies of about $10^4$ Hz- (i) an unsuccessful hopping occurs when an ion jumps back to its starting position and makes a correlated forward-backward-forward motion, and (ii) a successful hop occurs when the neighborhood ions relax in relation to the ion's position and stay in the new site. At high frequencies, there is a greater dispersive conductivity due to an increase in the ratio of successful to unsuccessful hopping. To comprehend better, the typical variation of ac conductivity with frequency at various temperatures is shown in **Figure 13**(a). For grain and grain boundary contributions in the polycrystalline ceramic materials, the ac conductivity will follow,[14,71,72]

$$\sigma_{tot}(\omega) = \sigma_{dc}(0) + A\omega^{n_1} + B\omega^{n_2} \dots (51)$$

where, A, B, $n_1$ and $n_2$ follows identical meaning like eq$^n$(50) for grain and grain boundary, respectively.

The thermal activation of charge carriers causes an increase in AC conductivity in a variety of materials, mostly semiconductors or dielectrics. Increased mobility and conductivity are made possible by more carriers gaining enough thermal energy to pass through potential barriers as the temperature rises. Thermal energy can also produce more electron-hole pairs in semiconductors by the impact ionization or thermal excitation and the conductivity rises with temperature due to these extra charge carriers. Enhanced conductivity also results from enhanced charge carrier scattering caused by more prominent lattice vibrations at higher temperatures. However, the variation of ac conductivity of metal with temperature is completely different and ac conductivity typically decreases with temperature. The lattice vibrations in metals, or phonons, intensify with increasing temperature, and the mobility of electrons is hindered by these phonons as they go through the metal lattice. With temperature, there is a drop in conductivity and an increase in resistance due to the increased electron-phonon scattering in metal. Because of increased phonon scattering, the mean free path of electrons (the average distance an electron travels between scattering events) decreases with temperature. Because of this, electrons have less time to move without being scattered, which lowers their total mobility and lowers conductivity.[73,74] One typical variation of ac conductivity ($\sigma$) with temperature for both metals and semiconductors is shown in **Figure 13**(b).

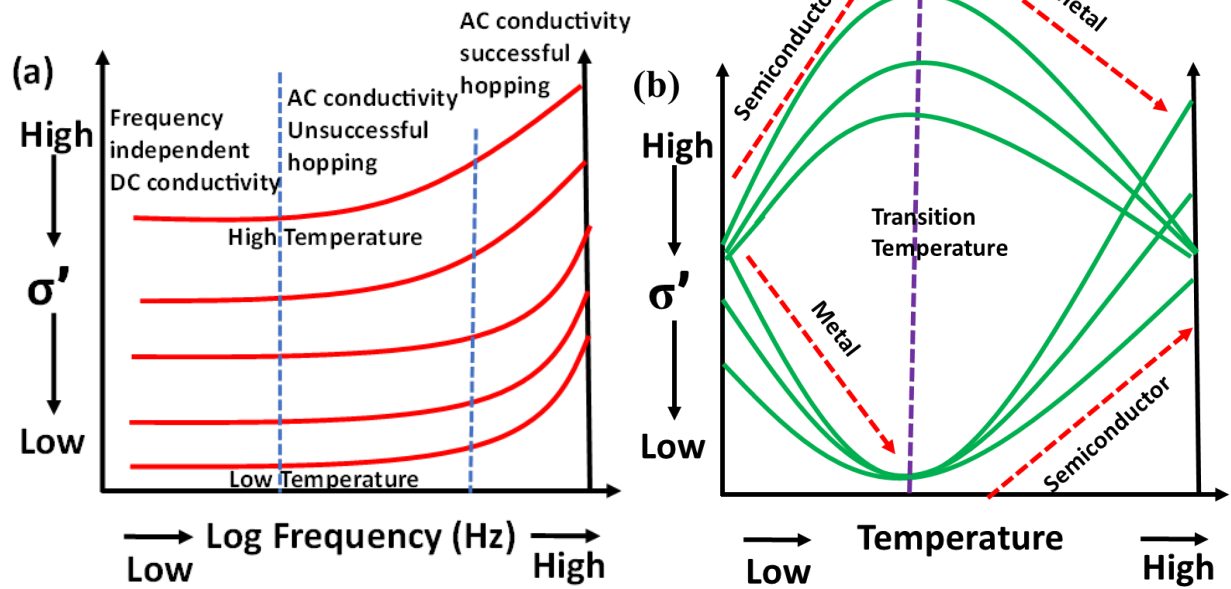

*Figure 13. (a) Typical plots of frequency dependence ac conductivity of a semiconductor/dielectric at various temperature, (b) The most frequent variation of ac conductivity with temperature at various frequency for metal and semiconductor.*

The variation of frequency exponent (*n*) with temperature plays an important role in ascertaining a particular conduction mechanism for the ac conductivity of any samples. Based on their fundamental conducting mechanisms-such as hopping conduction, ion migration, or electronic polarization, different materials have different frequency dependence with temperatures. Researchers ascertained that the conduction mechanism is more prevalent in a certain material by examining the frequency exponent and its value around 0 denotes Ohmic behavior or frequency-independent conductivity, and when its value over zero denotes frequency-dependent behavior, such as dispersive or capacitive effects. Numerous theories, including the overlapping large-polaron tunneling (OLPT), correlated barrier hopping (CBH), and quantum mechanical tunneling (QMT) models, have been put forth in the literature to explain the behavior of the exponents.[75] When the frequency exponent (*n*) is about equal to 0.8 and either grows slightly with temperature or is temperature independent, the quantum mechanical tunneling (QMT) model is the appropriate model to explain its conductivity. According to this model, the frequency exponent (n) can be expressed as,[76–78]

$$n = 1 - \frac{4}{\ln\left(\frac{1}{\omega\tau_0}\right)} \quad \ldots (52)$$

where $\tau_0$ is the characteristic relaxation times. From the eq$^n$(52), it can be observed that the frequency exponent (*n*) is temperature-independent but frequency-dependent. One can obtain also a temperature-dependent frequency exponent in the pair approximation within the context of the QMT model by considering the formation of non-overlapping small polaron by the charge carriers. This model is known as the non-overlapping small polaron tunneling model (NSPT) and the frequency exponent is given by[36,79,80]

$$n = 1 - \frac{4}{\ln\left(\frac{1}{\omega\tau_0}\right) - \frac{\omega_H}{k_B T}} \quad \ldots (53)$$

The term $\omega_H$ denotes the polaron hopping energy and the frequency exponent increases with the increase of temperature. Small polarons, which are localized distortions or deformations in the crystal lattice brought on by the movement of charge carriers (electrons or holes), constitute the foundation of this model. In particular, the tunneling of these tiny polarons between non-overlapping potential wells in the crystal lattice is taken into account by the NSPT model. The electron moves in a sequence of hopping between localized states in the non-overlapping tiny polaron tunneling model, and each hop is made possible by phonon emission or absorption. The temperature, the strength of the electron-phonon interaction, and the barrier height between states are some of the variables that affect the tunneling probability.

An additional theoretical model for explaining charge transport in materials, especially those with strong electron-phonon interactions is the overlapping large polaron tunneling (OLPT) model.[81–83] Large polarons include a more extensive deformation of the lattice surrounding the electron as opposed to small polarons. The electron can be successfully drawn through the material by this distortion. The overlapping model recognizes that there may be some degree of spatial overlap between the phonon wave function and the electron wave function, in contrast to the non-overlapping assumption of the small polaron model. The transport properties of materials may be considerably impacted by this overlap. The radius of a polaron is often significantly greater than the lattice constant of many materials and this polaron is referred to as a large polaron (also known as a Fröhlich polaron). The Fröhlich coupling constant ($\alpha$), which is a unitless quantity that widely characterizes the behavior of such a polaron,

$$\alpha = \frac{2\pi e^2}{hc}\sqrt{\frac{2\pi m_b c^2}{2\hbar\omega_{LO}}}\left(\frac{1}{\varepsilon_\infty} - \frac{1}{\varepsilon_0}\right) \ldots (54)$$

where $\varepsilon_\infty$, $\varepsilon_0$, $m_b$, and $\omega_{LO}$ represent high-frequency dielectric constant, low-frequency dielectric constant, effective mass, and longitudinal optical angular phonon frequency, respectively. The exponent ($n$) in the OLPT model is dependent on both frequency and temperature; it decreases to a minimum value as temperature rises and then rises as temperature rises. The expression for ac conductivity and frequency exponent according to the overlapping large polarons model can be described as,[84,85]

$$\sigma_{ac}(\omega) = \left[\frac{\pi^4 e^2 k_B^2 T^2 N^2}{12}\right] \times \frac{\omega R_\omega^4}{2\alpha k_B T + \frac{W_{HO} r_P}{R_\omega^2}} \ldots (55)$$

The $R_\omega$ is known as the hopping length and can be calculated by solving a quadratic equation,

$$R_\omega'^2 + [W_{HO} + Ln(\omega\tau_0)]R_\omega' - W_{HO}r_0' = 0;\ R_\omega' = 2\alpha R_\omega \text{ and } r_0' = 2\alpha r_0$$

According to the OLPT model by A.R. Long,[86] the polaron hopping energy ($W_H$) is reduced due to the overlap of two different sites and the large polaron wells and it is related to the activation energy ($W_{HO}$) related to the charge transfer between the overlapping sites by,

$$W_{HO} = \frac{W_H}{[1 - \frac{r_P}{R}]} = \frac{e^2}{4\varepsilon_P r_P}$$

Therefore, $R_\omega$ and frequency exponent (n) for the OLPT model can be determined by,

$$R_\omega = \frac{1}{4\alpha}\left\{\left(Ln\left(\frac{1}{\omega\tau_0}\right) - \frac{W_{HO}}{k_B T}\right) + \left[\left(\frac{W_{HO}}{k_B T} - Ln\left(\frac{1}{\omega\tau_0}\right)\right)^2 + \frac{8\alpha W_{HO} r_P}{k_B T}\right]^{1/2}\right\} \ldots (56)$$

$$\text{and, } n = 1 - \frac{8\alpha R_\omega + \frac{6W_{HO}r_P}{R_\omega k_B T}}{[2\alpha R_\omega + \frac{W_{HO}r_P}{R_\omega k_B T}]^2} \quad \ldots (57)$$

where, $r_P$, $\varepsilon_P$, $\alpha$, and N denote the polaron radius, effective dielectric constant, inverse localization length, and density of defect states, respectively.

When the frequency exponent (*n*) decreases with the temperature increases, the correlated barrier hopping (CBH) model is the appropriate model to explain its conduction mechanism. According to the CBH model, the conduction happens by a single-polaron or bi-polaron hopping process over the Coulomb barrier, which divides two defect centers. In this model, the frequency exponent (*n*) can be described as,[87–89]

$$n = 1 - \frac{6k_B T}{W_M - k_B T Ln(\frac{1}{\omega \tau_0})} \quad \ldots (58)$$

where $W_M$ is the binding energy which is the energy needed to transfer an electron entirely from one location to another and $\tau_0$ is the relaxation time with typical values ~$10^{-13}$ s. For large values of $W_M/k_B T$, the expression for *n* reduces to

$$n = 1 - \frac{6k_B T}{W_M} \quad \ldots (60)$$

As the $W_M$ decreases with increasing temperature it leads to decrement of frequency exponent as we can see from eq[n] (60). The typical variation of frequency exponent (*n*) with temperatures of QMT, NSPT, OLPT, and CBH model is shown in **Figure 14**.

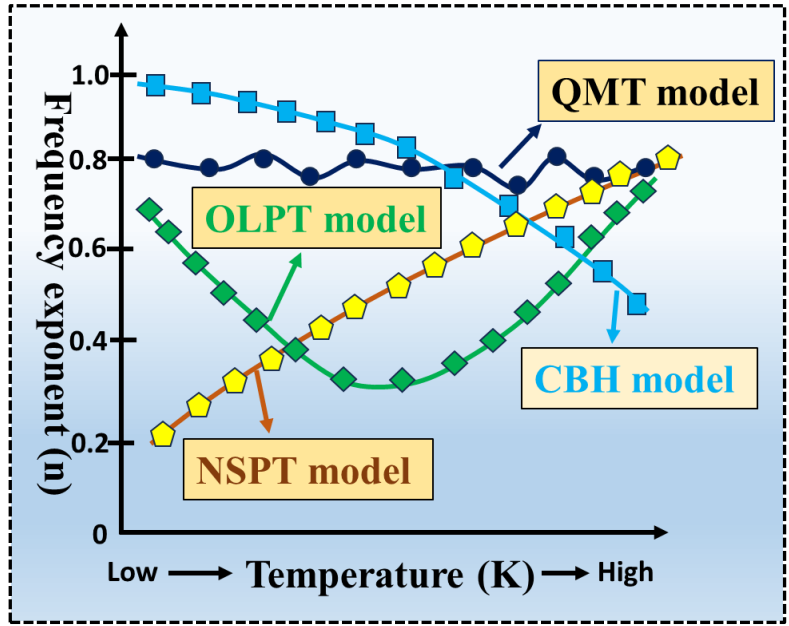

*Figure 14.* *The typical variations of frequency exponent (n) with temperature for four different conduction models.*

## 4. Impedance Spectroscopy for Electrochemical System

Electrochemical impedance spectroscopy (EIS) is widely used in many scientific and engineering domains because it can give deep insights into the electrical, kinetic, and transport characteristics of many electrochemical systems. A deeper comprehension of the basic electrochemical processes, i.e., charge transfer, ion transport, charge kinetics, etc. taking place within a system can be attained by researchers by examining the EIS spectra. The functionality of many electrochemical devices depends on the characteristics of the electrode/electrolyte interface, double layer capacitance, charge transfer resistance, and interfacial capacitance, and researchers can evaluate the performance of many electrochemical reactions from EIS spectra and identify the element that limits the reaction kinetics, surface coverage, and diffusion. Besides, the EIS spectra are used to study the corrosion process including corrosion rates, mechanism and effectiveness of coatings, battery/fuel cell performance, sensors and biosensors, and quality control and process monitoring.[90] A nice overview of the important application area of EIS spectra is schematically highlighted in **Figure 15**.

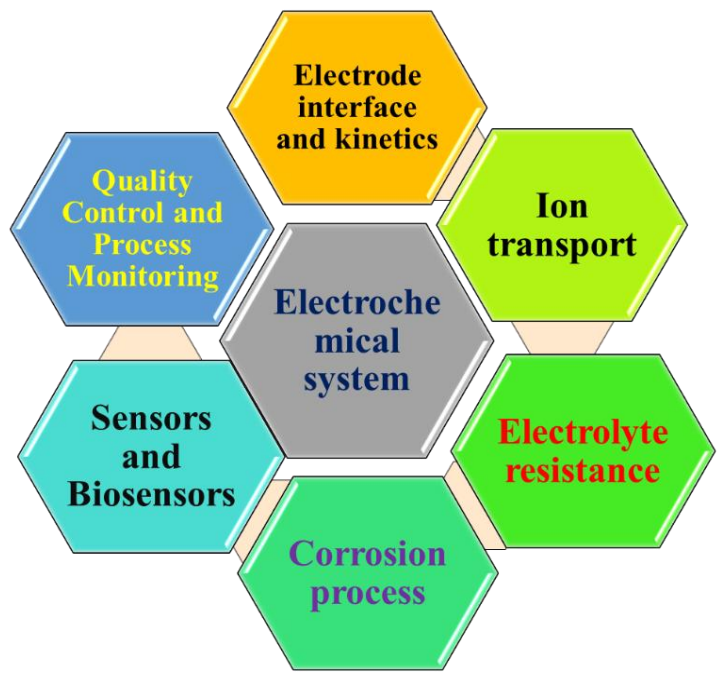

*Figure 15.* *Some important applications areas of EIS spectra in various fields of science and engineering.*

## 4.1. Frequency response analyzer (FRA) and electrochemical cell

An equipment specifically designed to measure the frequency response of electrical and electronic systems is called a frequency response analyzer, or FRA. Its broad range of applications in industries including electronics, telecommunications, power systems, materials research, and electrochemistry are facilitated by its deep insights into system behavior. An EIS experiment entails measuring the impedance of the cell under investigation following its excitation with a small-amplitude sinusoidal perturbation signal (for instance, voltage) at multiple excitation frequencies, typically within the range of a few mHz to 1 MHz. While excitation frequencies as high as 100 kHz are adequate for many applications, measurements up to a few MHz are necessary for applications involving solid electrolytes to fully resolve the impedimetric behavior at the high-frequency range. Typically, the sinusoidal perturbation signal is placed on top of a direct current (dc) signal, like the formal potential of a redox probe or the open circuit potential (OCP), which is the point at which the cell's current is zero. **Figure 16** shows a simplified schematic and working principle of a potentiostat/galvanostat-FRA equipment for an EIS experiment. By comparing the sinusoidal response, in this example $i(t)$, with two synchronous reference signals-one in-phase $[sin(\omega t)]$ and the other out-of-phase by 90° $[cos(\omega t)]$-with the sinusoidal voltage perturbation, FRA establishes the impedance of the cell under investigation. The resultant waveform is multiplied by the cell response, i(t), which is supplied into both correlators. The real and imaginary components of the impedance are then determined when the signals are driven to the integrators, which successfully eliminate all harmonic responses except from the fundamental. For more details, readers can follow these references. The real ($Z'$) and imaginary parts ($Z''$) of impedance can be obtained by,[91–94]

$$Z'(\omega) = \frac{1}{N_T} \int_0^{N_T} i(t) \sin(\omega t)\, dt \ldots (61)$$

$$Z''(\omega) = \frac{1}{N_T} \int_0^{N_T} i(t) \cos(\omega t)\, dt \ldots (62)$$

where $N_T$ is the number of periods that the signal integration is carried out. When the number of periods grows, the integration of the signal leads to a more efficient removal of noise from the response signal at the cost of an extended measurement time, which may enhance the stability of the electrochemical system. FRA is frequently used in electrochemistry to investigate the electrical characteristics of sensors, batteries, fuel cells, electrochemical cells, and corrosion processes through impedance spectroscopy. It is capable of describing phenomena like diffusion impedance, electric double-layer capacitance (EDLC), equivalent series resistance ($R_s$), and charge transfer resistance ($R_{ct}$).

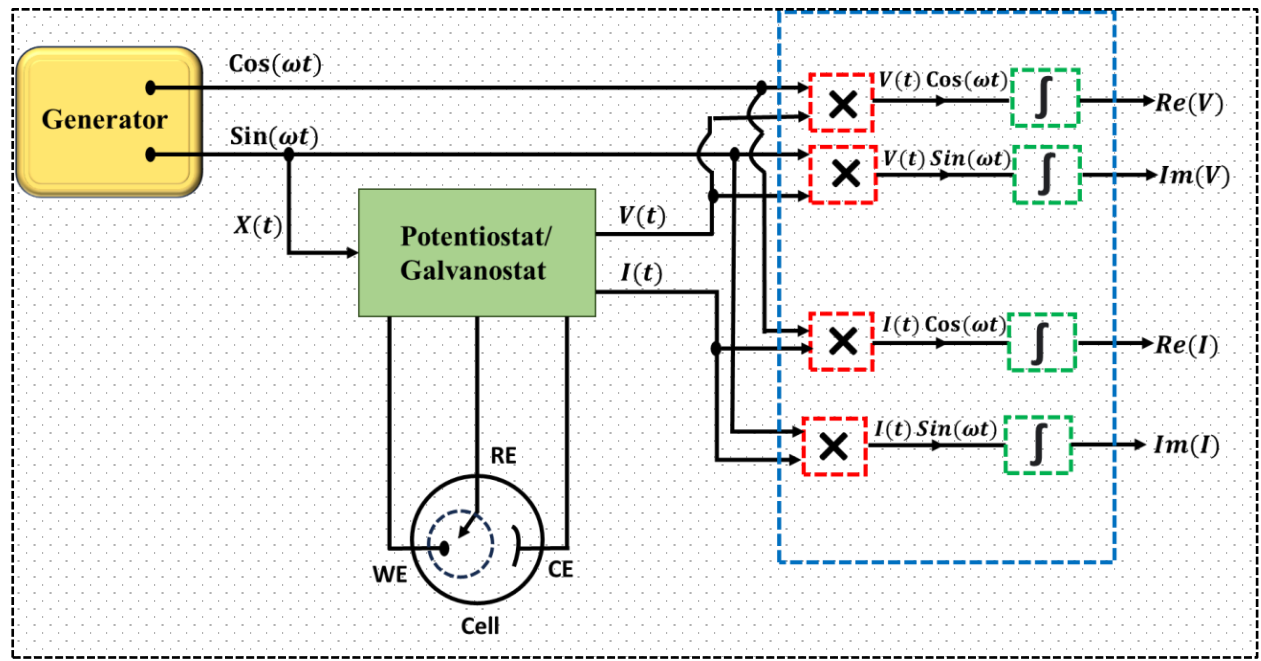

*Figure 16: Schematic process of working principle of a frequency response analyzer using a potentiostatic/galvanostatic EIS system.*

There are three different electrochemical set-ups are available to test the electrochemical performance-

*(i)    Three-electrode cell*: A three-electrode cell is one of the most common electrochemical cells which is comprised of three electrodes: a working electrode, a reference electrode, and a counter electrode, one schematic is shown in **Figure 17**a. An external power source or potentiostat regulates the potential difference in a three-electrode cell between the working and reference electrodes.[95–97] The electrical circuit is completed by the counter electrode, which permits the current to flow and maintains the intended potential difference between the working and reference electrodes. Typically, inert materials like gold, platinum, or graphite are used to make counter electrodes. In an electrochemical experiment, the potential of the working electrode varies while the resulting current is monitored. This makes it possible for scientists to investigate a range of electrochemical processes, including corrosion, electrodeposition, oxidation-reduction reactions, and electrocatalysis. A steady reference point for electrochemical measurements is provided by the reference electrode, which guarantees the accuracy and repeatability of the potential obtained at the working electrode. The impedance to the current flow in this configuration under a small amplitude voltage perturbation is caused by (i) the uncompensated/solution resistance (Rs), or the ohmic resistance of the electrolyte between the reference and the working electrode. Rs is determined by the separation between the reference and working electrodes for a particular electrolyte. In reality, this ohmic resistance also includes the resistance of the working electrode and the connection cables, which are typically ignorable. The charging and discharging of the electric double layer at the electrode/electrolyte interface in an alternating current environment behave like a capacitor and is represented by the symbol $C_{dl}$. The definition of the polarization or contact resistance $R_{ct}$ at steady-state observations is the slope of the voltage/current curve $R_{ct} = \Delta V/\Delta i$, and the steady-state conditions in EIS are roughly estimated as the frequency (f→0) approaches zero. The details of Warburg resistance ($Z_W$) have been discussed later.

*(ii)* *Two-electrode cell:* In comparison to a three-electrode cell, a two-electrode cell, sometimes referred to as a two-electrode setup, is a more straightforward electrochemical setup and there are only two primary parts- a working electrode and a reference electrode (**Figure 17**b).[97–99] An external power supply or potentiostat regulates the potential difference between the working electrode and the reference electrode in a two-electrode configuration. Current flows between the working electrode and the reference electrode as a result of the electrochemical reaction at the working electrode with the electrolytes. The potential of the working electrode is adjusted, and the resulting current is measured. The potential of the working electrode is measured from a stable and well-known electrochemical potential provided by the reference electrode. Standard hydrogen electrode (SHE), silver/silver chloride electrode (Ag/AgCl), and saturated calomel electrode (SCE) are common reference electrodes used in two-electrode setups. Usually, the working electrode and the reference electrode are in direct contact with an electrolyte solution. In a two-electrode, two-terminal configuration, the voltage applied between the counter (auxiliary) electrode and the working electrode and the impedance to the current flow is caused by the ohmic resistance of the working solution and the $C_{dl}$ of each electrode, given that both electrodes are perfectly polarizable. An electrical circuit with $R_s$, $C_{dl}$, WE, and $C_{dl}$, CE linked in series can be modeled using $R_s$ and the impedance of the cell, given that the total capacitance ($C_{tot}$) is provided by,[100]

$$\frac{1}{C_{tot}} = \frac{1}{C_{dl,WE}} + \frac{1}{C_{dl,CE}} \quad \ldots (63)$$

*(iii)* *Four-electrode cell*: For accurate measurements in electrochemical research, a sophisticated set-up called a four-electrode electrochemical cell is employed and the electrode configuration comprises four elements: the working electrode, the counter electrode, the reference electrode, and the auxiliary electrode (WS) which is displayed in **Figure 17**c.[101–103] The function of working, counter, and reference electrodes are the same as three-electrode configurations as mentioned. The purpose of the auxiliary electrode, sometimes referred to as the guard electrode or compensating electrode, is to reduce the impact of polarization and solution resistance. It is placed in close proximity to the working electrode in order to minimize undesired potential drops in the solution and maintain a more consistent electric field. Usually, the auxiliary electrode is connected to the potentiostat or power supply terminal that is opposite the working electrode. Predominantly, the four-electrode cell offers a stable and accurate platform with sophisticated control over potential and current measurement for examining electrochemical events. An AC perturbation (galvanostatic measurements) is applied between the WE and CE in a 4-electrode (or 4-terminal) configuration (**Figure 17**c), while the potential difference between RE and WS, which results from the impedance of the active materials, is monitored. These measurements are frequently used to determine the conductivity of a solid electrolyte or thin polymer film.

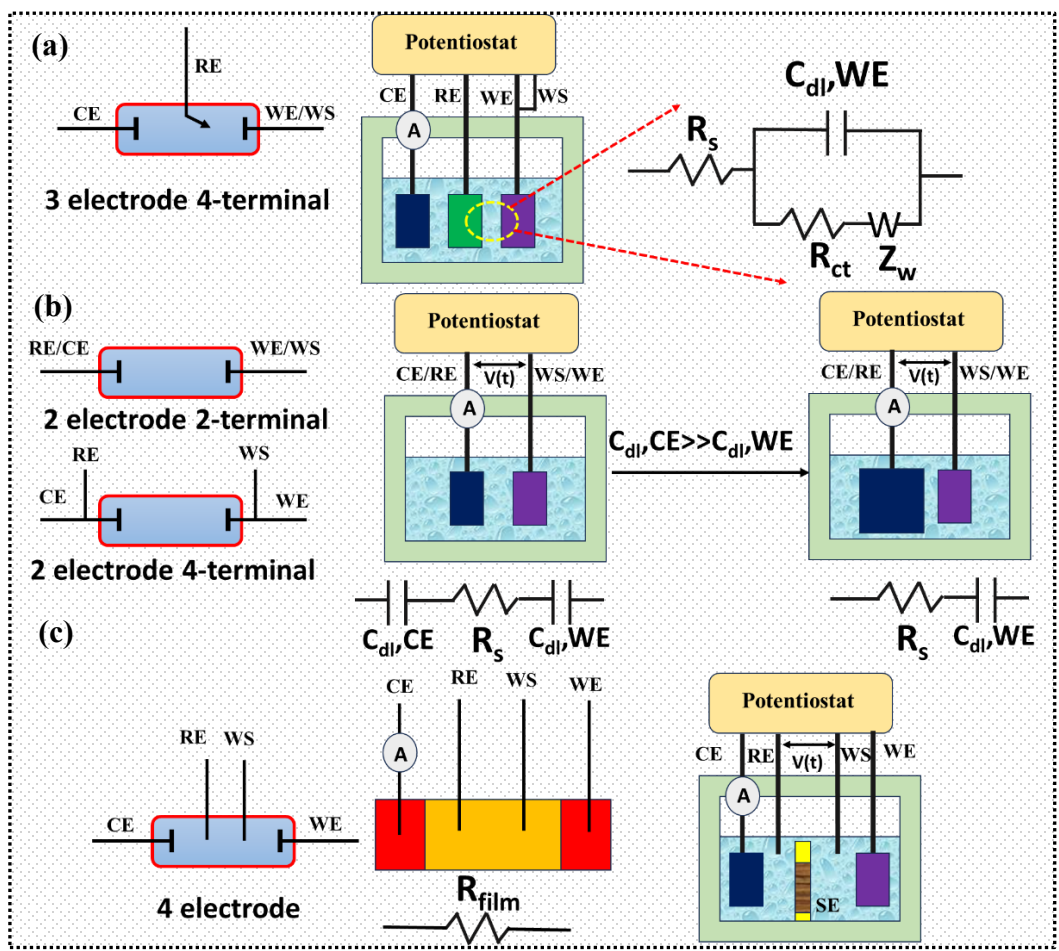

*Figure 17:* *Three different types of electrochemical cells of (a) three-electrode, (b) two-electrode, (c) four-electrode configurations with working electrode (WE), reference electrode (RE), counter electrode (CE), and working sense auxiliary electrode (WS) and their connections.*

### 4.2. Common equivalent circuit model:

Equivalent circuit models are frequently used in EIS to depict the electrical behavior of active electrochemical systems and these models are made up of various electrical circuits i.e. resistance (R), capacitance (C), inductance (L), Warburg resistance ($Z_w$), constant phase element (CPE) and their combinations that symbolize different electrochemical reactions taking place in the system. Here, we have discussed certain common components and circuits that are often seen in EIS analysis, even though the particular equivalent circuit model selected relies on the features of the system under investigation.[7,104,105]

(i) In a resistance (R) and capacitance (C) in series combinations, the voltage drop across the resistor ($V_R$) is proportional to the current flowing through it, and the voltage across the capacitor ($V_C$) is governed by the capacitance and the charge stored on its plates. The RC network always displayed a time-dependent behavior and the circuits involving impedance matching, time delays, and signal filtering frequently use this arrangement. The total impedance for this combination is,

$$Z = Z' + jZ'' = Z_R + Z_C = R + \frac{1}{j\omega C} \quad \ldots (64)$$

where, $Z' = R$ and $-Z'' = \frac{1}{\omega C}$

The typical Nyquist plots for the series RC network are shown in **Figure 18**a.

(ii) The LCR circuit comprised of inductors (L), capacitors (C), and resistors (R) in series is very significant in the electrical engineering and physics domain due to their ability to exhibit intricate electrical behavior. The LCR circuits are important in filtering and signal processing, frequency selectivity, impedance matching, sensor, resonant circuits, power factor corrections, etc. The impedance in the series LCR circuit is,[105]

$$Z = Z_R + Z_R + Z_C = R + i\omega L - \frac{i}{\omega C} = \sqrt{R^2 + (\omega L - \frac{1}{\omega C})^2} \quad \ldots (65)$$

where, $Z' = R$ and $Z'' = (\omega L - \frac{1}{\omega C})$

The Nyquist plots of the series LCR circuit are displayed in **Figure 18**b where the negative values come due to contributions of inductance. Parallel LCR circuits also exist where all three components are connected in parallel combinations.

(iii) In many electrical and electronic applications, resistor-capacitance coupled RC networks in parallel combinations are indispensable for tasks like impedance matching, voltage division, filtering, and timing. They are essential elements of contemporary engineering design and technology because of their simplicity and versatility. The overall impedance of the RC network is,[106]

$$\frac{1}{Z} = \frac{1}{Z_R} + \frac{1}{Z_C} = \frac{1}{R} + j\omega C \quad \ldots (66)$$

$$\text{or, } Z = \frac{R}{1 + Rj\omega C} = \frac{R}{1+(\omega RC)^2} - j\frac{\omega R^2 C}{1+(\omega RC)^2}$$

where, $Z' = \frac{R}{1+(\omega RC)^2}$ and, $-Z'' = \frac{\omega R^2 C}{1+(\omega RC)^2}$

canceling $\omega$ from both parts of impedance we can get,

$$(Z' - \frac{R}{2})^2 + Z''^2 = (\frac{R}{2})^2 \quad \ldots (67)$$

which can produce a perfect semicircular arc in the Nyquist plots shown in **Figure 18**c.

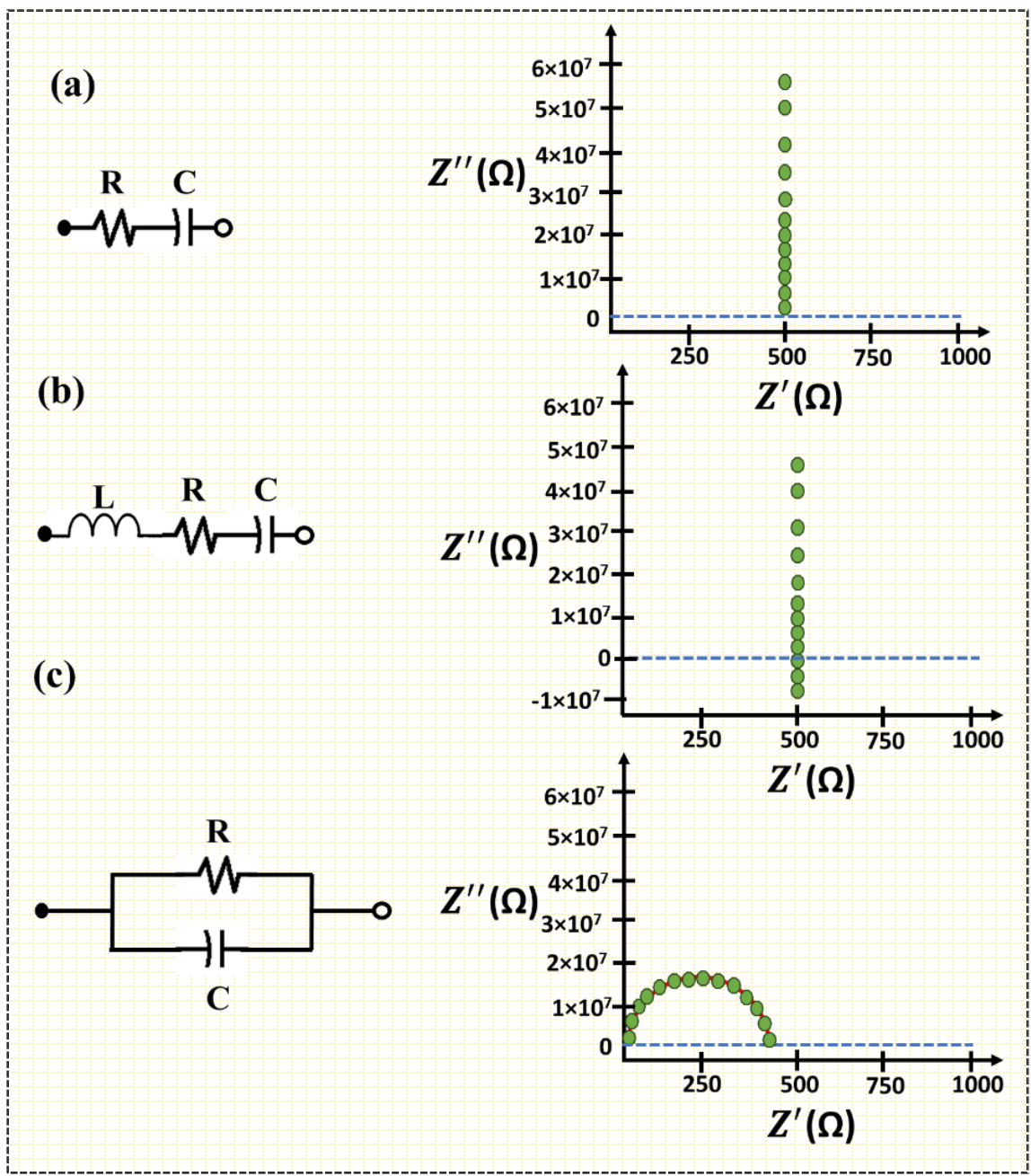

***Figure 18:*** *(a) The series RC circuit, (b) series LCR circuit, (c) parallel RC network, with their respective Nyquist plots.*

## 4.3. The Randles Circuits

In electrochemical impedance spectroscopy, the Randles circuit, named after the British electrochemist A. J. Randles, is a frequently used equivalent circuit model and it is employed to characterize the behavior of impedance in electrochemical systems, especially where the electrode reactions occur in solution and follows the faradaic reaction.[107–109] A variety of electrical components are combined to form the Randles circuit, such as solution resistance ($R_s$), charge transfer resistance ($R_{ct}$), electric double layer capacitance ($C_{dl}$), constant phase element (CPE), and Warburg resistance (W).

    (i) *Solution resistance ($R_s$):* The solution resistance ($R_s$) in the Randles circuit model denotes the resistance of the electrolyte solution that the current passes through it during electrochemical reactions. This is the first component of the circuit model that takes into consideration the ionic conductivity of the electrolyte as well as any resistance arising from the solution and cell connections. It is frequently shown with the other circuit elements in series with a basic resistor ($R_s$) and the Randles circuit model must include $R_s$ to appropriately reflect the impedance behavior of electrochemical systems. At high frequencies, the $R_s$ can be obtained from the intercept on the real axis ($Z'$) in Nyquist plots produced from EIS data. The high-frequency limit of the Nyquist plot can be used to determine its magnitude, which offers important details about the resistance of electrolyte solution and ionic conductivity.[104]

    (ii) *Charge transfer resistance ($R_{ct}$):* The charge transfer resistance ($R_{ct}$) in the Randles circuit model used in electrochemical impedance spectroscopy (EIS) denotes the resistance related to the charge transfer process at the electrode-electrolyte interface. Its value depends on numerous factors including the concentration of electroactive species in the electrolyte solution, the characteristics of the electrode surface, and the rate of electron transfer at the electrode interface. The $R_{ct}$ in the high-to-medium frequency range appears as a semicircular arc in Nyquist plots derived from EIS measurements. Larger semicircles correlate with lower $R_{ct}$ values because the diameter of this semicircle is inversely proportional to the value of $R_{ct}$. The intersection of the real and imaginary axes, which corresponds to the solution resistance ($R_s$), is usually represented by the center of the semicircle. The $R_{ct}$ can be obtained by fitting EIS data to the Randles circuit model and it offers important insights into the kinetics of the electrochemical reaction, the effectiveness of catalysts or electrode materials, and the efficiency of charge transfer processes. The following equation is used to determine the $R_{ct}$ values,

$$R_{ct} = \frac{RT}{k^0 n^2 F^2 AC} \quad \dots (68)$$

where, $k^0$, $n$, $F$, and $R$ represent heterogeneous electron transfer rates (cm s$^{-1}$), number of electron transfers during the electrochemical reaction, Faraday's constant with constant value 96485 C mol$^{-1}$, and universal gas constant (8.314 J mol$^{-1}$K$^{-1}$), respectively. The other two parameters $A$ and $C$ are associated with the active surface area of the working electrode (cm$^2$) and concentrations of the redox species, respectively.

(iii) *Electric double layer capacitance ($C_{dl}$):* The electric double layer capacitance ($C_{dl}$) is a capacitance related to the double layer formed at the electrode-electrolyte interface in the context of EIS and the Randles circuit model when an electrode is submerged in an electrolyte solution. The electric double layer is formed by a diffuse layer, which comprises mobile ions in the bulk electrolyte, and a Helmholtz layer, which contains selectively adsorbed ions next to the electrode surface. It represents the capacity of the double layer to store electrical charge and is determined by the surface area of the electrode, the composition of the electrolyte, and the potential applied to the electrode. The $C_{dl}$ in the Randles circuit model is commonly represented by a capacitor that is connected in parallel to the $R_{ct}$ and this parallel combination explains the capacitive behavior of the electrode-electrolyte contact and the charge transfer kinetics. The $C_{dl}$ is represented by a semicircular arc in Nyquist plots, and it usually overlaps with the $R_{ct}$ semicircle at the high-frequency region. The diameter of the semicircle indicates the magnitude of $C_{dl}$ and larger semicircles are associated with the higher capacitance values. The double layer has a thickness of one molecule, making it comparable to the dielectric layer in a traditional capacitor. When the capacitance is calculated using the early Helmholtz model, the model predicts a constant differential capacitance $C_d$ that is independent of the charge density but depends on the charge layer spacing δ and the dielectric constant ε by,[107]

$$C_d = \frac{\varepsilon}{4\pi\delta} \ldots (69)$$

(iv) *Constant phase element (CPE):* A common electrical element in impedance spectroscopy used to model non-ideal capacitive behavior is the constant phase element (CPE). [110–113] It is an extension of the traditional capacitor model and is especially helpful in explaining complex-behaving interfaces and systems with rough or porous surfaces. Most of the time, the impedance response of a system deviates from the ideal capacitor response and shows a -90° phase shift between voltage and current at all frequencies. The divergence in the impedance originates from assorted reasons i.e., non-uniform current distribution, electrode inhomogeneity, or surface roughness. If the semicircle is perfectly symmetric in that the diameter (equals $R_{ct}$) is twice the radius, $C_{dl}$ has been regarded as the perfect capacitor, with an impedance of

$$Z = \frac{1}{j\omega C_{dl}} \dots (70)$$

which is very uncommon. However, in non-ideal capacitive behaviors, the constant phase element is represented by Q and the impedance of the constant phase element ($Z_{CPE}$) is related to Q by,[27]

$$Z_{CPE} = \frac{Q}{(j\omega)^n} \dots (71)$$

The frequency-dependent phase angle of the impedance in the Nyquist plots is controlled by the CPE exponent (n). When n=1, the constant phase element behaves as an ideal capacitor, and n closer to 0, the CPE behaves as a resistor. For 0<n<1, the CPE indicates a more diffuse or non-uniform charge distribution by the impedance phase angle. **Figure 19**a shows the slanted/tilted impedimetric spectrum of $R_sQ_{CPE}$ circuits deviating from $R_sC_{dl}$ and **Figure 19**b shows the depressed semicircle of $R_s(Q_{CPE}R_{ct})$ circuit deviating from $Rs(C_{dl}R_{ct})$ due to the presence of a blocking electrode in the electrochemical cell. In both situations, the impedimetric data can be adequately described by using the so-called constant phase element (CPE) in place of an ideal capacitor ($C_{dl}$). The deviations of phase θ form the ideal phase (φ=90°) are related to the CPE exponent by, $\theta = (1 - n)90°$.[114,115] The variations of CPE with different n, φ, θ, and equations of impedance are shown in Table 1 with a schematic portrayed in **Figure 19**.

**Table 1:** *The modification of CPEs with various n values and their correlated impedance equations*

| Value of n | φ in (°) | Phase deviations (θ) in (°) | Equation of impedance |
|---|---|---|---|
| 0 | 90 | 0 | $Z_{CPE} = Q = R$ |
| 0.3 | 27 | 63 | $Z_{CPE} = \dfrac{Q}{(j\omega)^{0.3}}$ |
| 0.5 | 45 | 45 | $Z_{CPE} = \dfrac{Q}{\sqrt{(j\omega)}} = \dfrac{1}{Y_0}\sqrt{(j\omega)}$ |
| 0.7 | 63 | 27 | $Z_{CPE} = \dfrac{Q}{(j\omega)^{0.7}}$ |
| 0.9 | 81 | 9 | $Z_{CPE} = \dfrac{Q}{(j\omega)^{0.9}}$ |
| 1 | 0 | 90 | $Z_{CPE} = \dfrac{1}{j\omega C_{dl}}$ |

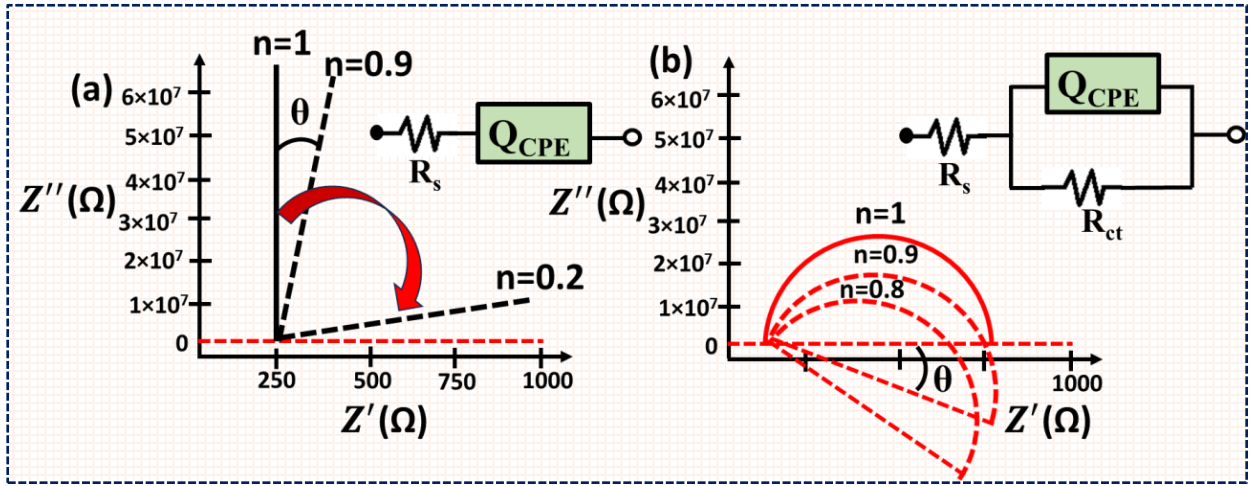

**Figure 19:** *The Nyquist plots of (a) $R_sQ_{CPE}$, (b) $R_s(Q_{CPE}R_{ct})$ circuit containing a constant phase element instead of $C_{dl}$.*

(v) *Warburg resistance (W):* Warburg impedance ($Z_W$) is an electrical element used in impedance spectroscopy to simulate the diffusion processes that take place in electrochemical systems.[6,107,116] It represents the impedance associated with the movement of ions or charge carriers through a solution or porous media towards an electrode interface. The Warburg impedance is especially important in systems like diffusion-controlled reactions at electrode surfaces where mass transport constraints are important. The impedance of the Warburg element can be obtained from **Table 1** for n=0.5 values. The Warburg impedance appears in a sloped line with a phase shift of -45° in Nyquist plots obtained from impedance measurements. Diffusion processes exhibit frequency-dependent behavior, as indicated by the slope of the line, which is inversely proportional to the square root of the frequency ($\sqrt{\omega}$). The Warburg impedance also represents the challenge of mass transport of the redox species to the electrode surface taking into account a semi-infinite linear diffusion. The series $R_W$-$C_W$ circuit in series has the same behavior as $Z_W$ where both $R_W$ and $C_W$ are frequency dependent and it can be expressed as[116]

$$Z_W = R_W + C_W = [\sigma\omega^{-\frac{1}{2}} - j(\sigma\omega^{-\frac{1}{2}})] \ldots (72)$$

$$\text{where, } \sigma = \frac{2RT}{n^2 F^2 C\sqrt{2D}}$$

The symbol D represents the diffusion coefficient ($cm^2 s^{-1}$) of the redox reaction considering the diffusion during oxidation and reduction reactions contributing equally.

**4.3.1. Simplified Randles Cell:** One of the most widely utilized models in the EIS is the simplified Randles Cell and it is shown in **Figure 20**a. It offers a simplified depiction of impedance behavior in any electrochemical system, especially when faradaic processes are present at electrode interfaces. Some of the primary areas where simplified Randles circuit is used are corrosion science, battery research, supercapacitors, fuel cells, sensors, biosensors, etc. It contains three components, $R_s$, $R_{ct}$, and $C_{dl}$ and the resultant impedance ($Z_{tot}$) will be[116,117]

$$Z_{tot} = R_s + \frac{1}{\frac{1}{R_{ct}} + j\omega C_{dl}} = R_s + \frac{R_{ct}}{1+(\omega C_{dl} R_{ct})^2} - j\frac{(\omega C_{dl} R_{ct})^2}{1+(\omega C_{dl} R_{ct})^2} \ldots (73)$$

Extracting the real and imaginary components of impedance we get,

$$Z' = R_s + \frac{R_{ct}}{1+(\omega C_{dl} R_{ct})^2}$$

$$\text{and, } -Z'' = \frac{\omega(C_{dl} R_{ct})^2}{1+(\omega C_{dl} R_{ct})^2}$$

canceling $\omega$ from both parts of impedance,

$$(Z' - R_s - \frac{R_{ct}}{2})^2 + Z''^2 = (\frac{R_{ct}}{2})^2$$

we can obtain a semicircular equation in the Nyquist plots. For $\omega \to \infty$, $Z' \to R_s$ and $\omega \to 0$, $Z' \to R_s + R_{ct}$ and $C_{dl} = \frac{1}{\omega R_{ct}}$ can be calculated from the peak of the semicircular arcs.

**4.3.2. Mixed kinetic and diffusion control:** An extension of the conventional Randles circuit used in electrochemical impedance spectroscopy is the mixed kinetic and diffusion control Randles circuit which is used to simulate electrochemical systems in which the overall impedance behavior is strongly influenced by mass transport (diffusion) processes as well as charge transfer kinetics. Apart from the $R_{ct}$ and $C_{dl}$ found in the regular Randles circuit, additional components are included to account for the diffusion processes, and it is called Warburg impedance ($Z_W$). It reflects the impedance brought about by ions or other charge carriers diffusing into or out of the electrode surface and its value can be extracted by modeling the Nyquist plots having a sloping impedance element in the low-frequency regions, as illustrated in **Figure 20**b. Considering the electrochemical cell is controlled by both charge transfer and mass transfer (diffusion) processes, the overall impedance will be,[104,109]

$$Z = R_s + \frac{1}{j\omega C_{dl} + \frac{1}{R_{ct} + Z_W}} \quad \dots (74)$$

where the Warburg impedance $Z_W = \sigma\omega^{-\frac{1}{2}}(1-j)$ and

$$\sigma = \frac{RT}{n^2 F^2 C\sqrt{2}}\left(\frac{1}{C^*_{OXD}\sqrt{D_{OXD}}} + \frac{1}{C^*_{RED}\sqrt{D_{RED}}}\right)$$

Inserting the value of $Z_W$ in the eq$^n$(74) and solving we can get,

$$Z' = R_s + \frac{R_{ct} + \sigma\omega^{-\frac{1}{2}}}{(C_{dl}\sigma\omega^{\frac{1}{2}} + 1)^2 + \omega^2 C_{dl}^2(R_{ct} + \sigma\omega^{-\frac{1}{2}})^2}$$

and

$$Z'' = -\frac{\omega C_{dl}(R_{ct} + \sigma\omega^{-\frac{1}{2}})^2 + \sigma\omega^{-\frac{1}{2}}(C_{dl}\sigma\omega^{\frac{1}{2}} + 1)}{(C_{dl}\sigma\omega^{\frac{1}{2}} + 1)^2 + \omega^2 C_{dl}^2(R_{ct} + \sigma\omega^{-\frac{1}{2}})^2}$$

At low frequencies, $\omega \rightarrow 0$, the reactance of $C_{dl}$ will be extremely high and $X_{C_{dl}} \gg R_{ct} + Z_W$. Then $Z'$ and $Z''$ will be reduced to

$$Z' = R_s + R_{ct} + \sigma\omega^{-\frac{1}{2}}$$

$$Z'' = -\sigma\omega^{-\frac{1}{2}} - 2\sigma^2 C_{dl}$$

Eliminating $\sigma\omega^{-\frac{1}{2}}$ from both impedances, we can get a relation between $Z'$ and $Z''$

$$Z' + Z'' - R_s - R_{ct} + 2\sigma^2 C_{dl} = 0 \dots (75)$$

The low-frequency limit is represented by a straight line in **Figure 20**b with slope 1 ($\varphi = 45°$) when $-Z'' = 0$ (low-frequency range) is imposed in eq$^n$(75),

$$Z' = R_s + R_{ct} - 2\sigma^2 C_{dl}$$

At high frequency, $\omega \to \infty$, $i_{C_{dl}}$ becomes significant and $R_{ct} \gg Z_W$. The real and imaginary impedance will be,

$$Z' = R_s + \frac{R_{ct}}{1 + \omega^2 C_{dl}^2 R_{ct}^2}$$

$$Z'' = -\frac{\omega C_{dl} R_{ct}^2}{1 + \omega^2 C_{dl}^2 R_{ct}^2}$$

which is the same as the charge transfer simple Randles circuit equation and exhibits semicircular arcs in **Figure 20**b in the high-frequency region.

4.3.3. **Two Randles circuits in series:** Two Randles circuits coupled in series produce an analogous circuit model which is more complicated and may account for several electrochemical processes taking place in a system in a sequential manner. When two different electrochemical interfaces or reactions are present, each of which is characterized by a unique set of mass transport phenomena and charge transfer kinetics. Researchers can extract parameters related to the mass transport characteristics, charge transfer kinetics, and overall electrochemical behavior of the system with multiple interfaces or reactions by fitting the equivalent circuit model to the impedance data obtained from electrochemical measurements. This enables a deeper comprehension of intricate electrochemical systems and procedures. Depending upon the nature of interfaces and electrochemical reactions, the diameter of semicircular arcs can vary in two Randles circuits in series and the equivalent circuit with Nyquist plots is demonstrated in **Figure 20**c.

4.3.4. **Complex Randles circuit:** An analysis of the impedance response of an organic coating on a metal substrate in contact with an electrolyte can be conducted using a Randles circuit model to reflect the electrochemical activity of the system. It is important to remember that the conventional Randles circuit model needs to be expanded or changed in order to take into consideration the unique features and complexities of organic coating systems and their interactions with the metal substrate and electrolyte. In those systems, the Warburg impedance of simple Randles circuits will be replaced with other Randles circuits and the equivalent circuit with Nyquist plots can be visualized in **Figure 20**d.

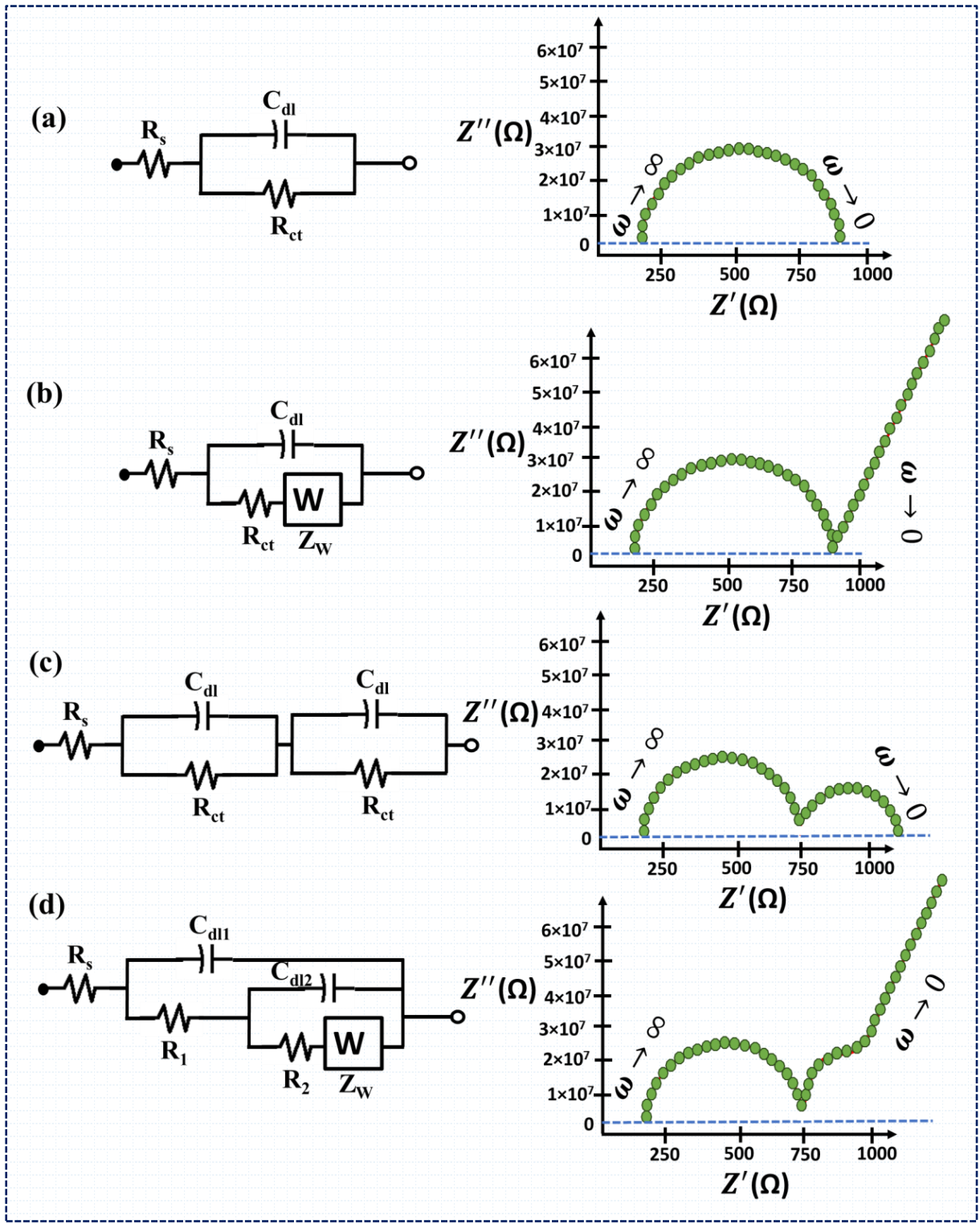

***Figure 20:*** *The typical equivalent circuit model with Nyquist plots of (a) a simple Randles circuit $R_s\,(C_{dl}R_{ct})$, (b) mixed kinetic and diffusion control Randles circuit with Warburg impedance ($Z_W$), (c) two Randles circuits in series, and (d) complex Randles circuits.*

**4.3.5. The transmissive and reflective boundary in the Randles circuits:** In practical scenarios, diffusion is frequently not semi-infinite. Such finite-length linear diffusion is noticeable in various applications such as conducting polymers deposited onto a planar electrode, in the internal diffusion into mercury film through the electrode, hydrogen diffusion into Pd or other hydrogen-absorbing material thin films or membranes, or in a rotating disk electrode where the diffusion layer matches the layer thickness. Two examples of finite-length diffusion are observable-transmissive boundary diffusion and reflective boundary diffusion.

The transmissive boundary diffusion occurs when a species penetrates one layer to the other layers along the x-axis and a steady state concentration gradient arises in the membrane. When hydrogen ions are reduced at the entry plane and hydrogen is oxidized and exits the membrane at the exit plane, transmissive boundary diffusion can be seen. The impedance in this kind of diffusion can be modeled by using the equation,[107,118]

$$Z_o(\omega) = \frac{1}{Y_o\sqrt{j\omega}} \tanh(B\sqrt{j\omega}) \quad \ldots (76)$$

where $Y_o$ is a parameter (Ohm$^{-1}\sqrt{s}$) that depends on the diffusion process and $B$ depends on the thickness ($\delta$ in cm) of diffusion layers and diffusion coefficient (D in cm$^2$s$^{-1}$) by,

$$B = \frac{\delta}{\sqrt{D}} \quad \ldots (77)$$

We have found that the diffusing layer thickness in rotating disc voltammetry (RDE) or with a transmissive boundary in solid oxide fuel cells or polymer membrane, is inversely related to the rotational speed by the equation,[119,120]

$$\delta = \frac{1.612 D^{1/3} v^{1/6}}{\sqrt{\omega}} \quad \ldots (78)$$

It can be observed from **Figure 21**a that the pattern of the impedimetric spectrum at low frequencies is approaching that of Warburg impedance $Z_W$ for semi-infinite linear diffusion when the rotation speed of the RDE is very low (the diffusion layer is thick).[107]

A plane is referred to as impermeable and a reflective boundary diffusion can occur when there is no possibility of charge transfer at $x=l$. When conductive polymers are deposited on metallic surfaces or when Pd is placed on a metallic surface (Au, Pt), this type of situation arises where additional adsorption into the base metal is not feasible. Lithium diffusion in thin solid-state oxide films, and hydrogen diffusion into thin films of different materials that absorb hydrogen or mass transport of charge carriers at redox polymer film-coated electrodes are typical examples of limited diffusion with reflecting boundary. The impedance in reflective boundary diffusion ($Z_T$) can be modeled by the equation,[121,122]

$$Z_T(\omega) = \frac{1}{Y_o\sqrt{j\omega}}\coth(B\sqrt{j\omega}) \quad \ldots (79)$$

As shown in **Figure 21**a, at low frequencies the diffusion of charges within the (polymer or oxide) films result in a $Z_W$ with a unity slope, while at decreasing frequencies the phase angle shifts from 45° to (almost) 90° due to a (purely) capacitive behavior caused by the finite thickness of the film combined with the blocking properties of the impermeable interface. Consequently, in the case of the transmissive boundary diffusion, the symbols have the same meaning as previously mentioned and an example of a simulated Nyquist plot of the total impedance at rising $B$ values is presented in **Figure 21**b, which represents a reflecting boundary mass transfer impedance with different B values.

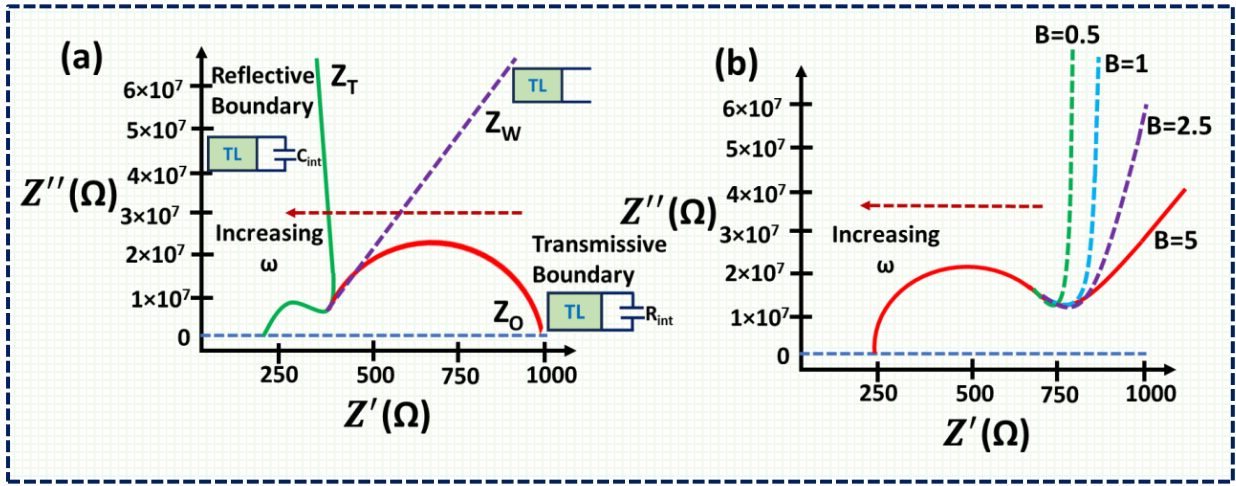

*Figure 21.* (a) Schematic illustrations of Nyquist plots at reflective and transmissive boundary mass transfer regime. (b)Typical Nyquist plots of reflective boundary diffusion with different B values from 0.5 to 5.

### 4.3.6. Inductive contributions in Nyquist Plots:

Deviations from the conventional semicircular arcs and linear tails in Randles circuits, another Nyquist plot with inductive contributions are observed by EIS spectroscopy. Usually, the inductive contributions are detected in the low-frequency region in Nyquist plots, and they can originate due to various effects, such as inductive effects related to the measurement setup, instrument response, and system internal properties. Inductive contributions to the Nyquist plots exhibit some extra features like spirals, loops, or non-linear trajectories in the low-frequency regime and these characteristics do not behave exactly as predicted by the Randles circuit model and another comparable circuits that are frequently used to analyze EIS data.[123–125] There are multiple reasons why inductive contributions may be seen in Nyquist plots and we have discussed the main three reasons here,

(i) The parasitic inductance found in the electrodes, cables, and electrical connections utilized in the impedance measurement setup can cause inductive effects, and the measured impedance exhibits phase changes and distortions due to these inductive components, deviating from the anticipated impedance response.

(ii) In response to the externally applied signal, the impedance analyzer device itself may display inductive behavior due to its internal circuitry, wiring, and component inductance, which can affect measurement outcomes and add to the inductive contributions seen in Nyquist plots.

(iii) In most cases, the system under study can display inductive behavior by nature because of eddy currents, magnetic fields, or other dynamic influences. Systems containing magnetic elements, ferromagnetic materials, or systems going through various dynamic processes can experience the inductive effects in the Nyquist plots.

In recent days, there has been a notable upsurge in spectra that display a "hook" or "loop" at the low-frequency region of the spectrum with a positive imaginary part in the Nyquist plot, as seen in **Figure 22**. This phenomenon appears consistently under steady-state conditions, allowing for the exclusion of measurement artifacts and these kinds of phenomena can be observed in lithium-ion batteries, proton exchange fuel cells (PEFC), organic light-emitting diodes (OLED), perovskite solar cells (PSC), thin film model electrodes, and the field of corrosion.[126] This characteristic has also been referred to as a (low-frequency) inductive loop, curl-back, or negative loop. The complex plane of the Nyquist plots in **Figure 22** can be split into two halves: a positive imaginary impedance half-plane (inductive) and a negative imaginary impedance half-plane (capacitive). It is customary to refer to features in the positive imaginary impedance half-plane as "inductive effects." Many researchers find it difficult to understand the low-frequency hooks in the Nyquist plots and describing the form of such an impedance curve by an equivalent circuit modeling is one of the big questions. The parallel connection of a resistor and a capacitor (RC circuit) and their equivalent circuit represents a complete semicircle of any dynamic system. **Figure 22**a represents a Nyquist plot containing a negative and positive imaginary impedance semicircle which can be modeled by $R_0(R_1C_1)(R_2C_2)$ and $R_0(R_1C_1(R_2L_2))$ equivalent circuit, respectively. The positive imaginary impedance is mainly associated with the inductance circuit in the low-frequency region and the diameter of semicircular arcs increases with the potential applied during electrochemical measurements.[127,128]

The fact is that both circuits frequently cannot account for a non-ideal semicircle for a practical system, and it manifests as a flattened or suppressed semicircle in both positive and negative imaginary impedance half-plane of the Nyquist plot. Nonetheless, real measurement data typically exhibit non-ideal semicircles for several reasons, including inhomogeneities, and the capacitor or inductor in those circuits is replaced by a constant phase element. The schematic plots of non-ideal semicircles containing CPE are shown in **Figure 22**b and the formula of CPE can be seen from eq$^n$(71). The CPE for inductance (L) and capacitance will be considered as, $Q = L^{-1}$ and $Q = C$, respectively. There are other different equivalent circuit models available for different processes mentioned above and the inductive contributions always exhibit positive imaginary impedance and a spiral shape in the low frequency region.

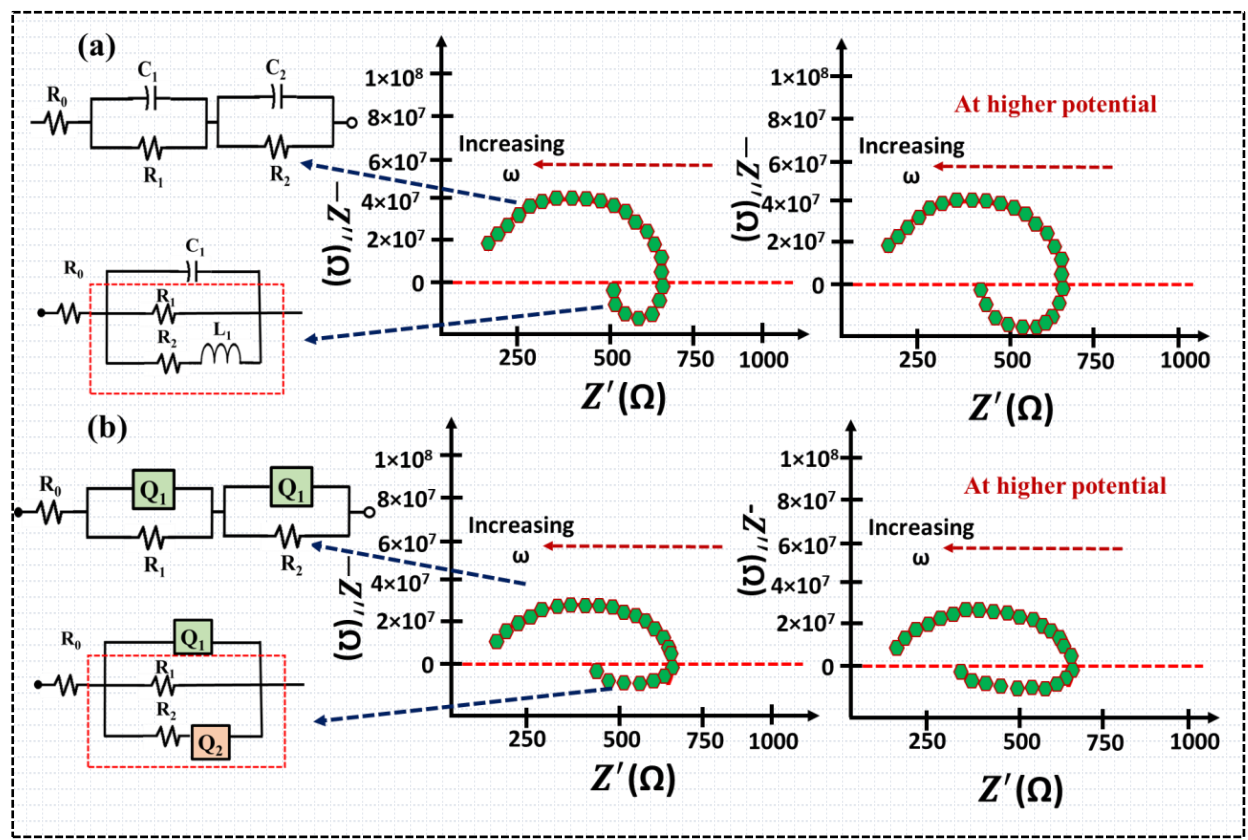

*Figure 22. (a) Typical Nyquist plots of one positive semicircle due to resistance and capacitance containing equivalent network, and negative semicircle due to inductance containing equivalent*

*circuit. (b) The Nyquist plots of one positive semicircle due to resistance and CPE (capacitance) containing an equivalent network, and a negative semicircle due to CPE (inductance) containing an equivalent circuit. At higher potential, the diameter of the semicircle in both cases increases.*

## 4.4. Applications of EIS spectroscopy

Electrochemical impedance spectroscopy provides valuable insights into the electrochemical behavior of materials and systems, it is used in a wide range of fields in energy sectors and bio-sensors. We have discussed here some key applications of EIS spectroscopy with their typical Nyquist plots-

### 4.4.1. Lithium-Ion Batteries

The evaluation and study of lithium-ion batteries (LIBs) are frequently investigated by EIS spectroscopy because of their capacity to offer insightful information about battery performance, electrode kinetic sand charge transfer process, degradation mechanisms, and overall health. One typical Nyquist plot of LIBs is shown in **Figure 23**. Three different regions are identified in the Nyquist plot of commercial LIBs: the low-frequency region is associated with the diffusion impedance, the high- and midfrequency region is associated with the capacitive reactance, and the superhigh-frequency region is affiliated with the inductive reactance. To fit this Nyquist plot, the second-order Randles circuit is typically utilized containing inductance, capacitance, and resistance. The inductive signal is observed at the superhigh-frequency region at the fourth quadrant, with a phase angle of ~ −90°. It is frequently seen in low-impedance systems, particularly in commercial LIBs. The phenomena are caused by spirally coiled electrodes, an outside lead, and the porous nature of electrodes. Sometimes, researchers eliminate the data of the inductive portion which is not a good idea because it can lead to a fitting error while fitting with the model, as mentioned in **Figure 23**. The solid electrolyte interface (SEI) and charge-transfer reactions of the cathode and anode are attributed to capacitive reactance and associated with the semi-arcs in the high- and mid-frequency region. The Warburg element is typically used to model the ion diffusion process in the solid phase, which is represented by the diffusion impedance at low frequencies.[104,129–131]

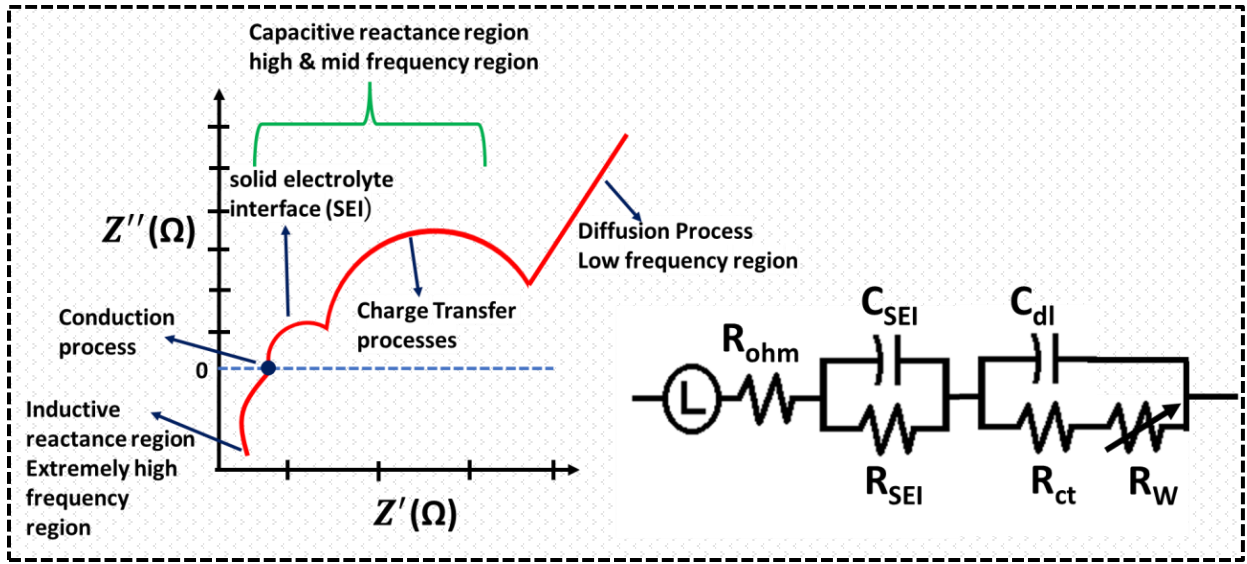

**Figure 23.** A typical Nyquist plot of LIBs with their dominant conduction processes at different frequency regions.

### 4.4.2. Solid oxide fuel cells

Solid Oxide Fuel Cells (SOFCs) are electrochemical devices that use high temperatures to reduce oxygen and oxidize fuel to directly transform chemical energy into electrical energy. Usually, SOFCs are hot devices because they operate at very high temperatures between 500°C and 1000°C. High fuel conversion efficiency and power density are produced by this high working temperature, which also facilitates rapid ion movement and effective electrochemical processes. Numerous fuels, such as hydrogen ($H_2$), natural gas ($CH_4$), biogas, syngas (a blend of hydrogen and carbon monoxide), and hydrocarbons, can power SOFCs and enable to production of high electrical efficiency between 50% to 60%. Solid oxide fuel cells (SOFCs) are galvanic cells that operate similarly to batteries, and they are comprised of two thin, porous electrodes, which are deposited on metal collectors and separated by a solid electrolyte. The typical Nyquist plot of SOFCs is shown in **Figure 24** with its equivalent circuit model. The Z-View software offers a complex nonlinear least-squares (CNLS) algorithm that is usually used to deconvolve the impedimetric data into these five distinct semicircles as processes (P); $P_{3A}$, $P_{2A}$, $P_{2C}$, $P_{1A}$, and $P_{ref}$. The following physical processes are depicted in this spectrum (**Figure 24**): overall ohmic losses are represented by $R_o$; gas diffusion coupled with charge transfer reaction and ionic transport within the anode functional layer is represented by $R_{2A}/Q_{2A}$ and $R_{3A}/Q_{3A}$ (where Q stands for CPE); The activation polarization of the cathode is represented by $R_{2C}$, a Gerischer element; the gas diffusion of $H_2/H_2O$ within the anode substrate is represented by $R_{1A}$, a generalized finite length Warburg element; and the reformate operation of the fuel cell is represented by $R_{ref}/Q_{ref}$. More details can be found in ref 132–135.

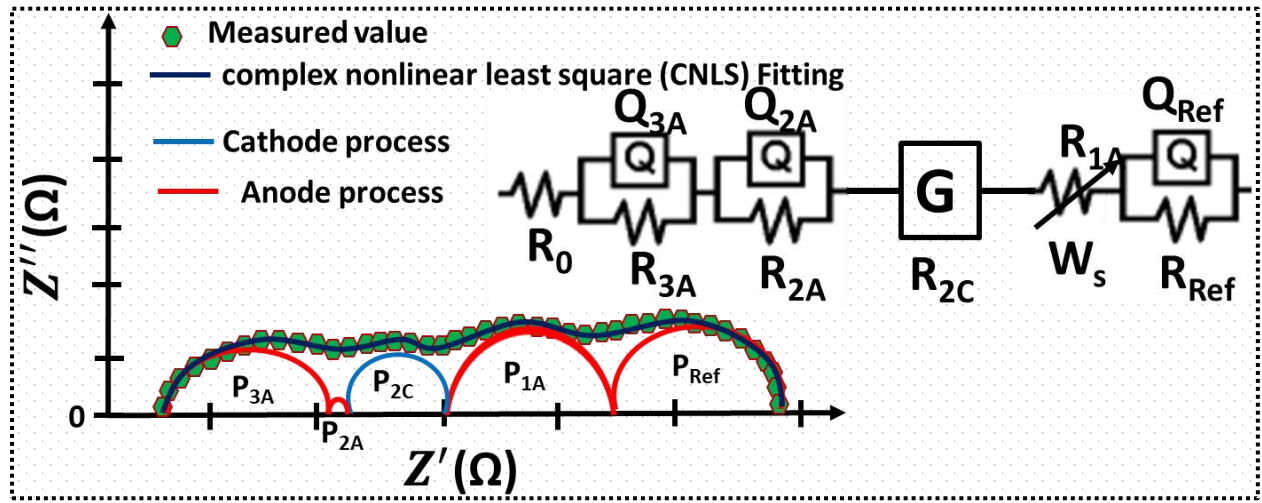

**Figure 24.** Typical Nyquist plot of SOFCs with CNLS fitting of impedance data.

### 4.4.3. Dye-Sensitized Solar Cells

One kind of thin-film photovoltaic device that turns sunlight into energy and works according to photoelectrochemistry principles, is called a dye-sensitized solar cell (DSSC), sometimes referred to as a Grätzel cell after the name of its inventor, Professor Michael Grätzel. Compared to conventional silicon-based solar cells, DSSCs have several benefits, such as flexibility, cost savings during production, and the capacity to produce energy in low light. The DSSCs are comprised of mainly 4 components i.e., semiconductor photoanodes, dye sensitizers, electrolytes, and counter electrodes. A light-absorbing dye, such as an organic or complex dye based on ruthenium, is usually applied on porous titanium dioxide ($TiO_2$) nanoparticles to create the photoanode. By absorbing photons, the dye molecules produce excited electrons that are pumped into the $TiO_2$ semiconductor when sunlight strikes them. Electric current is then produced by the electrons as they pass through an external circuit. The typical Nyquist plots of a DSSC are shown in **Figure 25** which relies on series resistance ($R_s$), contact resistance ($R_1$ or $R_{ct}$), CPE (1), recombination resistance ($R_2$ or $R_{rec}$), and CPE (2). The total resistance encountered by the current when passing through the different parts of the solar cell, such as the electrolyte, contacts, and internal parts of the semiconductor and conductive substrate, is referred to as series resistance ($R_s$). The resistance that charge carriers experience when they move across interfaces in a solar cell, such as the contact between an electrolyte and dye, or between an electrolyte and electrode, is referred to as charge transfer resistance ($R_{ct}$). The recombination resistance ($R_{rec}$) describes the ability of a solar cell to prevent charge carriers, such as electrons and holes from recombination which results in a reduction in efficiency and a loss of photocurrent when charge carriers come into contact with flaws or traps in the semiconductor material.[136–138]

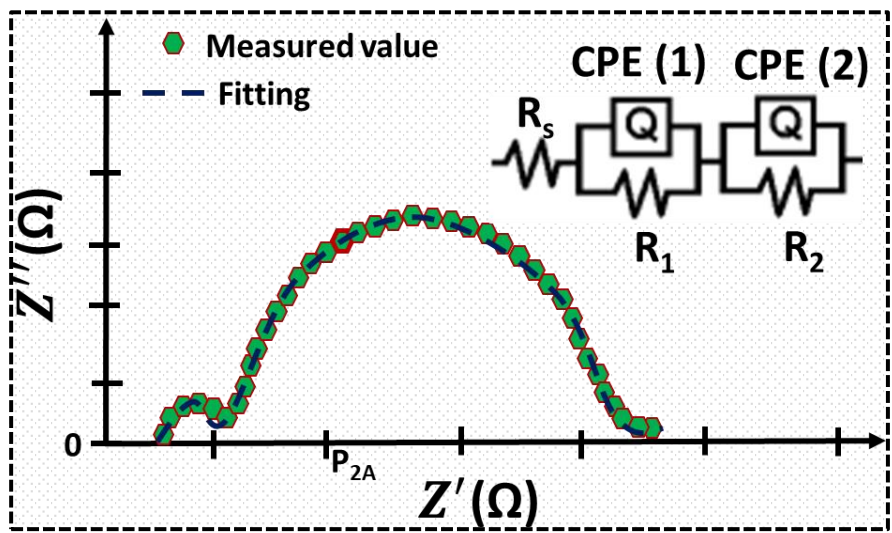

**Figure 25.** Nyquist plots of typical solid-state DSSCs with its equivalent circuit model.

### 4.4.4. Electrochemical impedimetric biosensors

The electrochemical impedimetric biosensors are analytical tools for the detection of biomolecules, including proteins, nucleic acids, and tiny molecules by monitoring changes in impedance at the electrode-solution interface. They work on the principle that target biomolecules change the electrical characteristics of the electrode surface, which results in variations in impedance. A biosensor is made up of a recognition element that recognizes one or more molecules in the material under investigation and subsequently, a variety of transducers (colorimetric, optical, electrochemical, or mass change) are employed to detect the recognition event. These transducers gather particular signals and amplify them to analyze the data. Biosensors have become indispensable diagnostic instruments because of their tiny sample requirement, good selectivity, reproducibility, quick detection, and high sensitivity. Remember that specific sensor configurations and sensing algorithms must be established for each target (analyte). According to the source, electrochemical biosensors are disposable, simple, inexpensive, portable, and easy to use and it is possible to perform electrochemical sensing by a standard three-electrode electrochemical cell which is made up of a working, counter (CE), and a reference electrode (RE). The Randles equivalent circuits which are usually used to fit the Nyquist plots of an impedimetric biosensor, as shown in **Figure 26**, simplify the resistance of solution ($R_s$), double layer capacitance at the electrode surface ($C_{dl}$), charge transfer resistance ($R_{ct}$), and Warburg resistance ($Z_w$). Diffusion taking place at the electrode-electrolyte contact produces Warburg resistance. Since the ideal capacitance behavior is not consistently observed in experiments, a constant phase element (CPE) is included to simulate or represent this non-ideal capacitance behavior. Faradaic impedimetric biosensors are typically considered to be more sensitive than capacitive biosensors.[9,139–141]

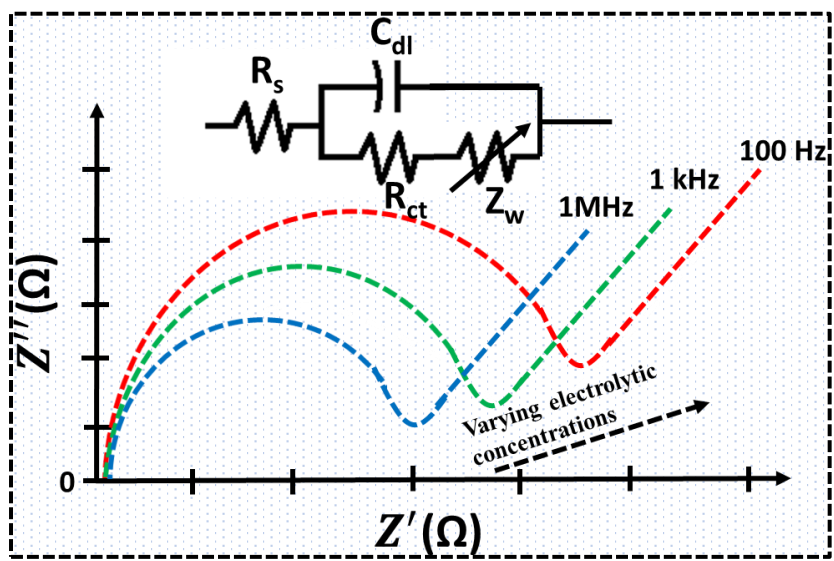

**Figure 26.** The typical Nyquist plots of impedimetric biosensors at various applied frequencies from 100 Hz to 1MHz.

## 5. Conclusions

In this review paper, we have mainly focused on the theory, principles, and applications of impedance spectroscopy of various electroceramic and electrochemical systems. Recently, electroceramic and electrochemical materials have been extensively investigated due to their wide range of applications such as electronics, energy storage, sensing, and catalysis and impedance spectroscopy is one of the effective analytical methods to determine their electrical and electrochemical properties. We have focused on the working principle, Nyquist plots with their equivalent circuit models, theory, and applications of impedance spectroscopy for both electroceramic and electrochemical systems. The detailed information on the electrical properties of electroceramic by measuring the complex impedance (real and imaginary parts) over a range of frequencies using impedance spectroscopy have been discussed in detail and aids in determining the dielectric constant, dielectric loss, and conductivity for their use in capacitors and other dielectric devices. The contributions of grain, grain boundary, electrode interface effect, Debye, non-Debye relaxation process, interfacial polarization, dielectric relaxation, and ac conduction mechanism, dielectric loss have been illustrated for various electroceramic systems with frequency and temperature evolution. The basic principle of electrochemical impedance spectroscopy and frequency response analyzer using a potentiostat/galvanostat has been demonstrated for two, three, and four-electrode cell configurations. Through the extensive measurement of impedance on any electrochemical system across several frequencies, EIS offers a comprehensive and refined comprehension of diverse mechanisms, including diffusion, series, charge transfer resistance, and double-layer capacitance. This method is very useful for analyzing and improving materials and equipment in the domains of electroceramics, fuel cells, batteries, sensors, and corrosion science. EIS is a potent tool for performance diagnosis and identification of degradation mechanisms due to its sensitivity to both fast and slow phenomena and its capacity to discriminate between various electrochemical processes. EIS is essential to the creation and advancement of sophisticated electrochemical systems because it provides information about the microstructural and interfacial characteristics of materials. It is one of the fundamental components of contemporary electrochemical research and development because of the improved material design, increased device performance, and extended lifespan of electrochemical technologies that result from its application.